# Late infall of molecular cloud material reshaped the outer Solar System

Elishevah van Kooten[1], Sebastian Sjørring Lodal[1], Isaac Onyett[2], Lasse Rasmus Pohlmann[1], Jean Bollard[1], Courtney Rundhaug[1], Martin Schiller[1], Martin Bizzarro[1]

[1] Centre for Star and Planet Formation, Globe Institute, University of Copenhagen, Øster Voldgade 5-7, 1350 Copenhagen, Denmark

[2] School of Earth Sciences, University of Bristol, Wills Memorial Building, Queen's Road, Clifton BS8 1RJ, UK

* Elishevah.vankooten@sund.ku.dk

## Abstract

Understanding the physicochemical evolution of the outer protoplanetary disk is critical, as it governed the distribution and delivery of key volatiles—such as water and organic compounds—to the inner, initially hot and volatile-poor terrestrial planet-forming region. These materials are essential for establishing potentially habitable environments and directly influence the emergence of life on rocky planets. Here, we extend beyond the traditional building blocks of the outer disk, the carbonaceous chondrite groups, and examine their ungrouped counterparts—the anomalous chondrites—to constrain a coherent model of disk evolution using Si, Mg, Fe, and Cr nucleosynthetic isotope systematics. Our results reveal that the outer disk was replenished via addition of isotopically distinct molecular cloud material that contributed a significant fraction of mass (>30%) to the building blocks of the gas giant accretion region. This late infalling material did not contribute to the main accretion phase of the terrestrial planets that have solely evolved their volatile budget from Ivuna-type planetesimals like Ryugu and Bennu, which, in our model, were formed at the inward-migrating water ice line. The perceived accretion of these icy planetesimals in the inner disk constitutes a fundamental shift in our perspective on the evolution of our Solar System.

## Main

Chondrites are minimally altered remnants of early Solar System asteroids that preserve a sedimentary record of dust and pebble accretion in the protoplanetary disk. Carbonaceous chondrites (CCs), which accreted beyond the snow line, exhibit diverse petrofabrics, varying widely in dust-to-pebble ratios and pebble sizes, where pebbles are represented mainly by chondrules that reflect once molten silicate spherules. Despite their significance, the origins of these textural differences remain poorly understood. For example, Ryugu- and Bennu-like bodies (CI-like chondrites) accreted predominantly fine-grained material, devoid of chondrules, while other CCs are dominated by (sub)millimeter-scale chondrules. Explaining such variation is essential for constraining mechanisms of mass transport and size sorting in the protoplanetary disk.

Understanding the formation of Earth and the terrestrial planets requires placing their accretion histories in the broader context of disk dynamics. Since Earth is thought to have incorporated a substantial fraction of material from beyond the snow line[1,2], a complete model of terrestrial planet formation depends on improved insights into the outer disk. Moreover, to effectively integrate cosmochemical data with astronomical observations and astrophysical models— particularly those probing outer disk regions (>5 AU) and features such as accretionary filaments and heterogeneous structures[3] —new cosmochemical constraints are needed.

Current models of the outer disk are anchored by two key observations. First, the moderately volatile element inventory of CCs reflects a dual-component mixture of volatile-depleted, once-molten chondrules and thermally unprocessed, solar-like matrix[4–6]. This is also reflected in the nucleosynthetic Cr isotope compositions of chondrules and matrix, where the decreasing chondrule/matrix ratio is correlated to enrichments in $^{54}$Cr[6]. Second, from a nucleosynthetic isotope perspective, CCs represent mixtures of two distinct dust reservoirs: CI-like material, chemically closest to the solar photosphere[7], and cometary dust associated with late infall of envelope material to the outer disk[8].

CCs are traditionally divided into eight major groups—CI, CM, CO, CV, CK, CR, CB, and CH—based on petrology and oxygen isotope systematics[9]. Bencubbin-like (CB) and CH chondrites, metal-rich and impact-generated[10], are thought to have formed at the time of disk dissipation. Hence, our understanding of the gaseous outer disk remains limited by the small variety of CC samples that together may not allow us to probe the entire outer disk. Recently, a growing number of anomalous CCs have been identified that do not fit within these traditional groups. These have been provisionally categorized (e.g., CT, CZ, CY, CL) based on oxygen isotope signatures[11,12], while others share isotopic affinities with known groups but display unusual chondrule/matrix ratios or chondrule size distributions[13]. These anomalous chondrites are vital for expanding our spatial and temporal understanding of the outer disk and for testing models of disk dynamics.

Here, we present Si, Mg, Fe, and Cr nucleosynthetic isotope data for fourteen anomalous and six grouped carbonaceous chondrites, combined with petrographic constraints on their petrofabrics (***Table 1; Supplementary Table 1; Supplementary Figs. 1–2***). These data constrain the chemical and physical evolution of the outer disk, the extent of radial mass transport, and the links between cosmochemical records, astrophysical models, and disk observations.

## Mixing two dust components in the outer disk

The recent identification of outer disk dark clasts (ODDs) in CR chondrites showed that carbonaceous chondrites define a Fe–Si isotope mixing array between CI-like dust and ODD-like, likely cometary, material[8]. The anomalous carbonaceous chondrites analysed here reinforce this view (***Fig. 1A***). Because the $\mu^{54}$Fe-$\mu^{30}$Si array is unrelated to chondrule/matrix ratios or other component abundances (***Table 1***; ***Supplementary Fig. 3***), it most likely reflects mixing between two precursor dust populations. A similar trend is observed in Mg-Si isotope space (***Fig. 1B***), where CI chondrites represent the $^{26}$Mg* (decay product of $^{26}$Al) and $^{30}$Si-rich endmember and ODD-dust the $^{26}$Mg* and $^{30}$Si-poor endmember. The $\mu^{26}$Mg* compositions reflect values corrected to a solar $^{27}$Al/$^{24}$Mg ratio (***Supplementary Fig. 4-5***), since we assume super-solar $^{27}$Al/$^{24}$Mg ratios represent the addition of CAIs to the bulk chondrite, which skew the bulk compositions to higher values, making comparison between chondrites difficult. Variations in the $\mu^{26}$Mg* values are attributable to either Mg isotope variations[14] or heterogeneity of the initial $^{26}$Al/$^{27}$Al ratio of the precursor dust[15–17]. If reflecting the former, we should expect to see correlations where the increase of $\mu^{30}$Si is at least an order of magnitude larger than that of $\mu^{26}$Mg* [18,19]. Since supernova nucleosynthesis models predict much larger relative overabundances of neutron-rich Si isotopes than Mg isotopes, any corresponding $\mu^{30}$Si effect would naturally be at least an order of magnitude greater than the $\mu^{26}$Mg* variations[18]. In contrast, the outer disk mixing line has a slope of 0.83 ± 0.17 (2σ, IsoplotR, model-1 regression), implying that $^{26}$Al heterogeneity is responsible for the linear trend between $\mu^{30}$Si and $\mu^{26}$Mg*.

Collectively, our data indicate a spatiotemporal isotopic gradient in the outer disk between CI- and ODD-like dust. Both endmembers are chemically CI-like but nucleosynthetically distinct, inconsistent with models in which the outer disk reflects mixing between non-carbonaceous and CAI-like isotope endmembers[20–22]. Chronological constraints remain limited, although similar CR–CI arrays occur in Fe–Ni isotope space, including CC-related iron meteorites[23]. Given modeled accretion times of <2 Myr for these irons and 2–4 Myr after CAI formation for carbonaceous chondrites[24,25], this gradient may have persisted for much of the disk lifetime.

Alternatively, the outer disk may have started as an isotopically CI-like reservoir that was later polluted by ODD-like material. This scenario is consistent with the late accretion of comets and CR chondrites[26,27], the absence of ODD-like material in the inner disk (***Fig. 1***), and evidence that CI-like material contributed to the terrestrial planet reservoir[1,2,28]. It also aligns with thermal-processing models in which inner-disk isotope variations reflect unmixing of dust populations present in CI chondrites[16,29]. Together, these observations favor the establishment of the outer-disk gradient through addition of ODD-like material to a CI-like reservoir.

An important implication is that all carbonaceous chondrites incorporated substantial ODD-like material (>30%; ***Fig. 1***). If carbonaceous chondrites represent outer-planet building blocks, then the gas giants accreted a significant fraction of isotopically distinct, cometary-like material.

## The role of distinct nucleosynthetic carriers

Systematic nucleosynthetic variations among carbonaceous chondrites were first recognized using Cr isotopes[30]. Correlated $\mu^{54}$Cr and $\Delta^{17}$O compositions define an apparent CI–CO mixing line[31], seemingly at odds with the CI–ODD mixing array identified here from Si, Fe, and Mg isotopes (***Fig. 2A***). Because CI chondrites have the highest $\mu^{54}$Cr values[30], they are commonly interpreted as fragments of bodies that accreted near the comet-forming region. At face value, the outer Solar System could therefore be described by three endmembers: $^{54}$Cr-poor CC material, CI-like material, and ODD-like material (***Fig. 2B***). However, such a model is difficult to reconcile with the two-endmember Si-Fe-Mg isotope systematics (***Fig. 1***). A third endmember lying on the CI–ODD array would generate the observed scatter in $\mu^{54}$Cr along the Si–Fe mixing line, but it cannot represent inner-disk dust without measurably displacing the array toward the NC field. It would therefore require a distinct $^{54}$Cr-poor carrier or dust component specific to carbonaceous-chondrite chondrules.

Another way to reconcile these models is to consider a different carrier phase for $^{54}$Cr relative to $^{54}$Fe, $^{26}$Mg* and $^{30}$Si. In Fe-Cr isotope space (***Fig. 2B***), bulk non-carbonaceous and carbonaceous chondrites plot in separate fields and are roughly anti-correlated within these fields, where $\mu^{54}$Fe increases when $\mu^{54}$Cr decreases. ODDs and CR chondrites are offset from the CC slope. Anomalous carbonaceous chondrites fill the gap between ODDs/CR and classical CCs. Assuming, here, that Solar System materials are governed by the same endmember trichotomy as observed in ***Figure 1***, we can draw a triangle connecting these endmembers (***Fig. 2B***), where all Solar System materials fall within this space. We observe that the $\mu^{54}$Cr values of carbonaceous chondrites are shifted down from the CI-ODDs mixing line towards the ureilite endmember. This suggests that, unlike $^{54}$Fe, $^{54}$Cr is hosted in a dominant carrier phase that responds differently to physical processing, likely reflecting distinct thermal properties of the isotope carriers rather than chemical behaviour intrinsic to Cr.

Two main trends unique to the nucleosynthetic Cr isotope variations provide insight into the nature of the $^{54}$Cr carrier. First, $\mu^{54}$Cr values are correlated to the matrix mass fraction of bulk carbonaceous chondrites, showing mixing between a $^{54}$Cr-depleted non-matrix fraction and an isotopically CI-like matrix fraction[6] (see also ***Extended data Fig. 1***). Second, $\mu^{54}$Cr values of bulk carbonaceous chondrites are also correlated with their $\Delta^{17}$O signatures[30,31], where the $\Delta^{17}$O values reflect both the composition of the rock as well as the volatile component (i.e., water and organic molecules). As such, $^{54}$Cr isotope variation seems linked to volatile depletion of chondrules and/or their precursors.

This implies that the $^{54}$Cr carrier is hosted in a volatile component that is lost during chondrule formation and/or during thermal processing of volatiles in chondrule precursors[32]. This interpretation is consistent with previous observations of $^{54}$Cr carrier phases in carbonaceous chondrites, which show that the $^{54}$Cr carrier resides predominantly in the hydrated silicate fraction of pristine CCs, whereas $^{54}$Cr hosted in refractory phases accounts for less than 1% of the total $^{54}$Cr budget [30,33]. This process has recently been identified as the origin of $^{96}$Zr variations in carbonaceous chondrites, where step-leaching experiments show that hydrated matrix fractions record the most extreme $\mu^{96}$Zr anomalies [34]. These data indicate that supernova-derived Zr was incorporated as atomic species into interstellar ice mantles following dust destruction in the ISM, establishing a volatile carrier of nucleosynthetic anomalies prior to Solar System formation. We note that since $^{54}$Cr nucleosynthesis can occur in various supernova-related environments[35] – with $^{30}$Si and $^{26}$Al being produced in pre-supernova massive stars or from explosive H burning during type II supernova explosions[36] – it is possible that their carrier phases reflect distinct condensation pathways.

Interestingly, if $^{54}$Cr depletions indeed reflect thermal processing of a $^{54}$Cr-rich volatile carrier, CI chondrites themselves may include an ice fraction that is reprocessed in the disk. Recent Cr isotope analyses have shown the existence of various Ryugu particles (i.e. C0002) with elevated $\mu^{54}$Cr values[37] relative to CI chondrites (***Fig. 3A***). Furthermore, studies of pristine, chemically CI-like dark clasts in polymict ureilites also have elevated $\mu^{54}$Cr values (***Fig. 3B***)[38–40]. Together, Ryugu particle C0002, ureilite dark clasts and ODDs fall on a linear correlation in $\mu^{53}$Cr versus $\mu^{54}$Cr space, which we interpret as the initial outer disk mixing line (***Fig. 3B***), where CI chondrites are $^{54}$Cr depleted relative to this line. As such, we propose that $^{54}$Cr-rich Ryugu particles (from here on: Ryugu*) represent the primordial disk composition, whereas CI chondrites reflect thermally processed Ryugu* material, where the $^{54}$Cr carrier has been (partially) lost from the ices (***Fig. 3C***). Carbonaceous chondrites represent variable mixtures of CI-/Ryugu*-like and ODD-like materials that have been thermally processed to different degrees.

## Mass transport from the inner to the outer disk

Our data imply that outer-disk evolution reflects mixing between cometary and Ryugu*-like dust, with subsequent thermal processing of the ice-hosted $^{54}$Cr carrier. This appears at odds with Cr-Ti-O isotope data for CV chondrules, which have been interpreted to reflect either inner–outer disk mixing[41] or transport of inner-disk chondrules to the CV accretion region[42,43].

Here, we argue that chondrule mass transport from the inner to the outer disk is limited to large CV chondrules similar in size range to CV CAIs and that the bulk of carbonaceous chondrite chondrules formed locally in the outer disk. Although the oxygen isotope composition of CV chondrules reflect that of outer disk chondrules[44], recent arguments supporting CV chondrule transport were made based on the inner disk $\mu^{54}$Cr signatures of chondrule cores ($\mu^{54}$Cr = –50 ppm) that were progressively enriched in $^{54}$Cr towards their fine-grained rims ($\mu^{54}$Cr = 125 ppm)[43], a trend which was later confirmed by Si isotopes[45]. Here, we combine the average Leoville CV3.1 chondrule core $\mu^{54}$Cr signature with that of its $\mu^{54}$Fe signature (***Fig. 4A***), confirming an inner disk origin of these chondrules. As such, we interpret the oxygen isotope compositions of CV chondrules to reflect the local disk environment in which they formed, where O isotope signatures falling below the terrestrial fractionation line are not unique to outer disk environments. The reported bulk $\mu^{54}$Cr of Leoville is 76 ppm, and using the average $\mu^{54}$Fe of literature CV bulk signatures[46], we find a cometary (ODD-like) signature for the composition of Leoville's matrix. Note that this mixing entirely excludes CI-like material in CV chondrites. In Si-Fe isotope space (***Fig. 4B***), we also plot the Leoville chondrule cores, but we have no constraints from the matrix composition. Future Fe isotope analyses of either Leoville bulk or matrix can verify the cometary nature of the Leoville fine-grained component. Nevertheless, based on the available data, it is evident that only the reduced CV carbonaceous chondrites have incorporated inner disk derived chondrules, since the bulk composition of these chondrites are expected to fall in the 'gap' between NC and CC in Si-Fe isotope space. As such, using the combined Si-Fe-Cr isotope systematics of chondrules, we propose that CV chondrites uniquely incorporated inner disk chondrules. These chondrules may be accessory to outward CAI transport by disk winds shortly after or during CAI formation when the disk was massive[47,48]. Alternatively, a more proximal accretion region for CV chondrites is envisioned, where CI-like dust was transported to the inner disk[43]. In both scenarios, we predict relatively old ages for CV chondrules with inner disk $\mu^{54}$Cr signatures. Note that CO, CM and CR chondrites also contain small and less abundant refractory inclusions that reflect outward transport of material, and it is unknown how this is translated to outward transport of equal sized chondrules.

Collectively, we suggest that most CC chondrules formed locally in the outer disk, whereas large reduced CV chondrules were transported outward with CAIs and accreted into CV chondrites. This is consistent with the chondrule size distributions of most samples studied here, which are similar to those of CM and CO chondrites (~300 μm; ***Table 1***) and likely representative of outer-disk chondrule formation.

### The mechanism of outer disk nucleosynthetic evolution

Our data suggest that the outer disk evolved from an isotopically Ryugu*-like reservoir toward a CI-like reservoir through two linked processes: addition of isotopically distinct dust by a molecular-cloud streamer[3,49]and thermal processing of $^{54}$Cr-rich ices in the disk.

Recent observations and models of young protoplanetary disks and their parent molecular clouds highlight the widespread presence of accretionary filaments, or streamers, that feed and replenish disks from large distances[49,50]. These streamers can deliver isotopically heterogeneous material from the broader giant molecular cloud, where nucleosynthetic products of supernovae and asymptotic giant branch stars are dispersed at random[51]. Their formation is linked to the filamentary structure of star-forming environments[3], and they are increasingly viewed as common features of protoplanetary disk evolution. Although their timescales, entry angles, and impact locations remain active areas of research, streamers can influence disk mass budgets, geometry, and chemistry, and may trigger instabilities that facilitate planetesimal formation[52]. A streamer impacting the outer Solar System could therefore introduce ODD-like nucleosynthetic material, process ice and dust through shock accretion[53], promote accretion of carbonaceous asteroids, and potentially contribute to gas giant formation.

Although our study does not directly constrain the timing of streamer arrival, meteorite accretion ages provide useful limits. Carbonaceous chondrites accreted within ~2.5 Myr of Solar System formation[24], placing an upper bound on the event. In Fe-Cr isotope space, non-carbonaceous chondrites and achondrites define a CI/Ryugu*-ureilite array that lacks ODD-like material, suggesting streamer input shortly before or after their accretion at <2 Myr after CAI formation[24]. Depending on outer-disk drift timescales, including ~$10^6$ yr for 100 μm particles drifting from 100 AU[54], non-carbonaceous bodies could accrete ODD-free while CC iron parent bodies accreted at similar times with an ODD-like contribution[23] (see also ***Supplementary Fig. 6***). Thus, the accretion ages and isotope systematics point to infall roughly around 2 Myr after CAI formation.

## The accretion regions of carbonaceous chondrites

The second process identified here is thermal processing of ices that shifted the $\mu^{54}Cr$ signatures of chondrule precursors. This could occur through vertical mixing of small grains to warmer disk surface layers, chondrule formation, or streamer-driven shock heating. The relative loss of $^{54}Cr$ from the initial Ryugu*-ODD mixing line may therefore reflect progressive loss of interstellar water ice, with CI chondrites representing Ryugu*-like material containing a processed ice component. Because this model requires processing of volatile rather than rocky material, we propose that CI chondrites accreted inside Jupiter's orbit, near the water ice line (***Fig. 5***). This location explains several key characteristics of CI chondrites:

1. Their scarcity of high-temperature refractory components such as chondrules and CAIs, likely filtered by a pressure barrier such as Jupiter[47]. Smaller fragments could pass through, as indicated by $^{16}O$-rich anhydrous silicates[55] and microchondrule remnants[56] in Ryugu particles.

2. Accretion near the water ice line would involve sublimation and recondensation of water ice, facilitating loss of nucleosynthetic isotopes trapped in ISM ices while accounting for abundant ice in CI chondrites.

3. The prevalence of CI-like bodies such as Ryugu and Bennu in the asteroid belt could be explained if these bodies formed rapidly via accretion at the water ice line[57].

4. The proposed inward flux of CI-like, in contrast to ODD-like pebbles, to the terrestrial planet region[1,2] is more plausible if CI chondrites accreted inside Jupiter's orbit or a comparable pressure barrier.

Recasting CI chondrites as inner Solar System materials requires a fundamental shift in assumptions about outer disk planetesimal provenance and protoplanetary disk evolution. The prevailing paradigm places CI chondrites at the largest heliocentric distances among carbonaceous meteorites[29,58], a view supported by samples from the CI-like asteroids Ryugu and Bennu. These samples contain complex organic inventories and isotopic signatures commonly interpreted as evidence for formation in cold, outer-disk environments[59,60]. For example, Bennu samples record high ammonia abundances implying accretion beyond the ammonia ice line[77] and organic phases consistent with pre-aqueous cryochemical processing of $NH_3$- and $CO_2$-rich ices[61], whereas Ryugu contains organic hotspots enriched in $^{15}N$ and deuterium[78]. Placing CI chondrites, including Bennu and Ryugu, near the water ice line therefore requires rethinking ice-line location and dynamics, material transport, and disk processing. Moreover, Bennu and Ryugu are heterogeneous rubble piles that sample different source reservoirs and degrees of parent-body processing[62,63], complicating their relationship to uniquely CI-like material.

Over the past decade, several xenolithic clasts hosted in meteorites have emerged as potential analogues of cometary material. The Zag clast, a centimeter-scale, organic-rich inclusion within the Zag meteorite (an H chondrite), exhibits nitrogen and carbon abundances comparable to particles from Bennu and ranks among the most organic-rich materials known[64]. Likewise, lithic clasts in the Isheyevo meteorite contain abundant $^{15}$N-rich organic matter interpreted to have accreted beyond the $NH_3$ and HCN ice lines, yielding some of the most $^{15}$N-enriched bulk signatures measured in meteoritic materials, approaching cometary values[65]. More recently, ODDs identified in the CR chondrite NWA 14250 exhibit organosulfide-rich regions resembling GEMS-like matter, adjacent to $^{15}$N- and D-rich organic domains, with minimal evidence of aqueous alteration[8]. The molecular structure of these organics closely resembles that of interplanetary dust particles related to comets (CP-IDPs[66]). Their bulk oxygen isotope compositions are among the most $^{16}$O-poor measured and overlap with the CP-IDP field. Collectively, the Zag clast, Isheyevo clasts, and ODDs define an "outer-disk mixing line" in N–H isotope space, reflecting mixing between interstellar medium–derived organic matter and $^{15}$N-rich ices[8]. Importantly, this mixing line extends to endmembers that are more $^{15}$N- and D-enriched than materials incorporated into canonical carbonaceous chondrites, which contain comparatively $^{15}$N-poor ices and isotopically lighter organic matter. Notably, bulk CI chondrites do not reach the $^{15}$N, D and $\Delta^{17}$O enrichments characteristic of comets and inferred cometary analogues described above. Nevertheless, even if CI chondrites no longer retain their long-held status as the Solar System's outermost, comet-like materials, their importance – and that of sample-return missions to CI-like asteroids such as Ryugu and Bennu – is only amplified, as they provide critical constraints on the delivery of volatile and prebiotic compounds that contributed to Earth's accretion and early chemical inventory[59,67,68].

In summary, we demonstrate that carbonaceous chondrites—including recently recognized grouplets and anomalous chondrites—represent mixtures of a primordial disk component, isotopically similar to $^{54}$Cr-rich Ryugu particles, and outer-disk dark clasts that likely reflect cometary material. We interpret the latter as evidence for late-infalling streamer pollution of the outer disk, which thermally altered the ISM-derived ice component carrying the anomalous $^{54}$Cr signature (***Fig. 5***). Chondrules that formed locally in the outer disk have incorporated this thermally processed material to varying degrees and are mixed with ODD-like matrix. Inward transport is limited to fine-grained isotopically CI-like dust, whereas outward transport is limited to CAIs and coeval chondrules as found in CV chondrites. These CV chondrites accreted at a barrier that has up until now been viewed as defining the dichotomy between non-carbonaceous and carbonaceous chondrites. We propose here that CI chondrites formed inward of this barrier, close to the water ice line. Thus, CI chondrite material could have contaminated the inner disk as the water ice line migrated inward[69]. Earth and the terrestrial planets therefore represent mixtures of ureilite-like and CI-like materials, with limited outermost-disk contribution. Cometary material only contributed significantly to Earth's surficial layers as modelled for our planet's atmosphere from noble gas isotope systematics[70].

## Materials and Methods

### *Materials*

Anomalous chondrite samples in this work include C2- and C3-ungrouped with potential links to CT (Telekoast-type) chondrites (NWA 16569, NWA16559, Chwichiya 002, NWA 14051, NWA 14139, NWA 16001), CM-anomalous (NWA 16554, NWA 1256, NWA 16000), a potential CY

(Dhofar 2046), CR-anomalous (NWA 14674), SAH00182 and Tarda. In addition, we studied two CL (Loongana-type) chondrites (NWA 033, NWA 13400), Ankazomena (CVox3), three CO3.0 chondrites (NWA 10916, NWA 11085, NWA 16476) and a CM2 chondrite (NWA 12561), since CO chondrites have not been analyzed in our multi-isotope space. The samples used in this study were acquired from meteorite dealer Luc Labenne and are now part of the holdings of the Centre for Star and Planet Formation. At present, these meteorites are not curated within a formal collection.

### ***Sample preparation and petrographic characterization***

<u>Bulk chondrites</u>

The anomalous and grouped chondrite handpieces with fusion crust in this study were cut into multiple 2 mm slabs using a diamond wire saw at the Center for Star and Planet Formation (Globe Institute, University of Copenhagen). Some of the smaller handpieces were mounted into epoxy before cutting, since their fragility caused them to break up into small pieces otherwise. The saw dust (~100-200 mg) was collected and ~200-500 mg pieces without fusion crust were broken off to be crushed and processed further for digestion and chemical purification in the clean lab.

The slabs were polished down to 0.3 µm grid size using first SiC and then $Al_2O_3$ sheets with MQ water. The slabs were subsequently imaged entirely in back scattered electron mode by Zeiss EVO 15 scanning electron microscope at the Centre for Star and Planet Formation (University of Copenhagen), using variable pressure mode and avoiding carbon coating (see ***Supplementary Fig. 2***). Finally, these images were processed with ImageJ to extract chondrule-matrix ratios (***Supplementary Fig. 7***). Sample descriptions are provided in the supplementary text and high-resolution BSE images are included as raw data files.

Fractions of ~100-200 mg of the saw dust and crushed sample pieces were digested in 3:1 7M HNO3: 28M HF mixtures inside Parr bombs for 3 days (1 day at 150 °C, 2 days at 210 °C). Afterwards, the solutions were dried down and taken up in aqua regia for 2 more days on hotplate. The aliquots were then processed through Fe, Mg and Cr purification techniques [43,46,71] and a 5% fraction of the sample solution was saved for iCAP inductively coupled plasma mass spectrometry to assess their weathering grade from their Sr and Rb anomalies (***Supplementary Fig. 8***, ***additional data table R1***).

Another 5% fraction was used to determine Al/Mg ratios by MC-ICPMS. In detail, a 6M HCl solution was loaded on a 200-400 mesh Eichrom AG1-X8 anion resin (1 ml resin volume) to separate Fe from the matrix. This step was repeated to ensure complete removal of Ni and Co from the Fe cut. The matrix cut was pretreated for 3 hours on a hotplate at 130 °C in 10M HCl to ensure speciation to Cr(VI)[72]. The sample was diluted to 0.5M HCl and loaded on a 1 ml AG50-X12 200-400 mesh cation resin to separate Cr (+Na, K) in 20 ml 0.5M $HNO_3$ from high field strength elements in 3ml 1M HF and Mg (+ Ni, Mn, Ca) in 10 ml 6M HCl. This step was repeated to ensure >99% separation of Cr and Mg. Subsequently, the Cr cut was dried down and taken up in 400 µl 1M $HNO_3$ and 100 µl concentrated $H_2O_2$. The samples were left at room temperature for >1 week to ensure complete speciation to Cr(III) and then diluted to 0.5M $HNO_3$. The sample was loaded onto a 1 ml AG50-X12 200-400 mesh cation resin to remove Na and K (9 ml 0.5M $HNO_3$) and trace amounts of Fe (3 ml 1M HF). The final Cr cut was eluted in 10 ml 6M HCl. The Mg cut was further purified to remove Ni, Mn and Ca according to techniques described in Bizzarro et al. [71]. BHVO2 and DTS-2b reference standards were processed alongside the samples (***raw data table R2***).

For the silicon (Si) chemistry analysis, a second portion of the bulk powder was processed using NaOH fusion and column purification methods, as outlined by Onyett et al.[16]. Specifically, the powdered samples were fused with NaOH in silver crucibles at 720 °C for 13 minutes. The resulting fusion cake was then dissolved in Milli-Q water and acidified with nitric acid ($HNO_3$). After dissolution, silicon was purified through cation exchange chromatography using columns packed with 3 mL of AG50-X12 resin (200–400 mesh). Approximately 1 mg of silicon was loaded onto each column for the samples, and around 300 µg for the quartz sand standard NBS-28. To reduce matrix effects caused by sulfate ions ($SO_4^{2-}$), which are not retained by the cation exchange resin, the solutions were spiked with a sulfate solution to maintain a fixed sulfur-to-silicon (S/Si) ratio of ~5.

_Individual chondrules_

Type I chondrules in CV chondrite Leoville were identified and imaged in BSE mode by Zeiss EVO 15 SEM (StarPlan) and subsequently excavated using a Dremel with tungsten carbide drill bits attached. Care was taken to avoid the surrounding matrix, which was aided by the sharp contrast in powder color (dark grey for matrix and yellow for chondrules). The powders were transferred to Savillex beakers in MQ water and digested using the same Parr bomb techniques as outlined above for the bulk chondrites. These chondrules were then processed through Fe and Cr chemistry using the purification methods described for the bulk chondrites, with the exception that the resin volumes were downsized by a factor 4 to accommodate the smaller sample sizes and maintain low blank levels.

### *Multi-collector inductively coupled plasma mass spectrometry*

The Fe, Cr, Mg and Si isotope dataset and obtained Al/Mg ratios can be found in the ***additional data table R2*** (individual analyses) and ***Supplementary Table 1***. The Fe and Cr isotope compositions were measured using the Thermo Fisher Neoma MC-ICP-MS at the Centre for Star and Planet Formation (University of Copenhagen) in the medium-resolution mode (M/ΔM > 6000 as defined by the peak edge width from 5 to 95% full peak height) to resolve gas-based interferences on the high mass side. Fe aliquots were introduced to the plasma source using an Elemental Scientific Inc. (ESI) Apex2HF sample introduction system with an Actively Cooled Membrane (ACM) unit attached and with an uptake of 30 µl min−1. The detailed iron isotope acquisition setup follows Schiller et al.[46], where $^{60}Ni$ and $^{53}Cr$ were measured along $^{54}Fe$, $^{56}Fe$, $^{57}Fe$, and $^{58}Fe$ to correct for direct isobaric interferences on $^{54}Fe$ and $^{58}Fe$ from $^{54}Cr$ and $^{58}Ni$, respectively. We report our data relative to the IRMM-014 Fe isotope standard in the µ-notation, where the reported data represent the mean and 2SE of 10 individual standard-bracketed sample analyses, each comprising 25 × 16.7 s of on-peak baseline measurement and 200 × 8.3 s of sample measurement. Samples were typically analyzed 2-4 times. Cr aliquots were introduced using an ESI Apex HF sample introduction system with ACM unit and an uptake rate of 30 µl $min^{-1}$. The samples were measured without the use of an auxiliary gas to the introduction system to reduce gas-based interferences, at low radiofrequency power and sample gas inflow, with a Jet and X cone. The isotopes $^{49}Ti$, $^{51}V$, and $^{56}Fe$ were analyzed alongside $^{50}Cr$, $^{52}Cr$, $^{53}Cr$, and $^{54}Cr$ to correct for isobaric interferences. Data are reported relative to the SRM979 Cr isotope standard in the µ-notation, where the reported data represents the mean and 2SE of 10 individual standard-bracketed sample analyses, each comprising 75 s of on-peak baseline measurement and 100 × 8.3 s of sample measurement.

Mg isotope compositions were measured using the Thermo Neoma MC-ICPMS in the medium-resolution mode (M/ΔM > 6000) using an ESI Apex Omega introduction system and an uptake rate of 30 µl min−1. The isotopes $^{24}Mg$, $^{25}Mg$ and $^{26}Mg$ were analyzed using $10^{11}$ Ω resistors with typical signal intensities of 80-90V. Mg isotope data are reported relative to the DTS-2b standard in the µ-notation, where the reported data represent the mean and 2SE of 10 individual standard-bracketed sample analyses, each comprising 25 × 16.7 s of on-peak baseline measurement and 100 × 16.7 s of sample measurement.

**Data availability**

All data are available in the main text, the supplementary materials, and in the Additional Data Tables 1 and 2.

**Acknowledgements**

We give special thanks to Dr. Adrien Houge for providing us with figure 5 of our manuscript. We also acknowledge the valuable comments provided by two anonymous referees that helped improve this manuscript. We thank Dante Lauretta for helpful discussions on the provenance of Bennu lithologies.

**Author contributions:** Conceptualization: E.v.K., Methodology: E.v.K., M.B., and M.S. Investigation: E.v.K., I.O., L.R.P., S.S.L., M.S., M.B., C.R., J.B., Writing—original draft: E.v.K. Writing—review and editing: E.v.K., M.B., and M.S. Resources: M.B., Funding acquisition: E.v.K., M.B., and M.S. Visualization: E.v.K. and M.B. Supervision: E.v.K. Data curation: E.v.K. Validation: E.v.K. Formal analysis: E.v.K., Project administration: E.v.K.

**Competing interests:** The authors declare that they have no competing interests.

| | Official | Potential | Abundances | | | | | Ch size | Affinity (O) | $\Delta^{17}O$ | |
|---|---|---|---|---|---|---|---|---|---|---|---|
| | | | *Ch* | *mx* | *RI* | *Metal* | *Sulfide* | | | | |
| NWA 033 | CL4 | | 72* | 21* | rare | 6 | <0.1 | 455 ± 349 | $CV_{red}$ | -3.96 | [1] |
| NWA 13400 | CL3.9 | | 76* | 17* | 1 | 6 | <0.1 | 477 | $CV_{red}$ | -4.01 | [1] |
| SAH 00182 | C3-ung | CV-CR | 65 | 20 | 1 | 10 | <0.1 | ~1000 | $CV_{red}/CV_{oxA}$ | -3.89 | [2] |
| NWA 16569 (1) | C3-ung | CT3 | 60 | 35 \| ~50 | sparse | none | 15 | 360 ± 280 | CT (low $\Delta^{17}O$) | -6.32 | [3] |
| NWA 16569 (2) | | | | | | | | | | | |
| NWA 16559 | C3-ung | CT3 | 64 | 30 \| ~50 | sparse | rare | 3.5 | 400 ± 270 | CT (low $\Delta^{17}O$) | -7.28 | [3] |
| NWA 16554 (1) | C3-ung | CM2T-an | ~50* | ~40 | <1 | none | 2.1 | 210 ± 80 | CM (ext) | -2.81 | [4] |
| NWA 16554 (2) | | | 32 | 54 | 1.2 | none | 1 | 150-300 | | | |
| Tarda | C2-ung | | 6.5 | 93.3 | 0.2 | 0.5 | 2 | 260 ± 110 | Sub CI-CY | -0.28 | [3] |
| Chwichiya 002 | C3.00-ung | CT3 | 25 | 74 | none | rare | <0.1 | 480 ± 300 | CT | -3.88 | [4] |
| NWA 14051 | C3-ung | CT3 | 31 | 72 | none | 0.5 | <0.1 | 210±160 | CT | -3.97 | [4] |
| NWA 14139 | C3-ung | CT3 | 25 | 73 \| *~50* | none | 0.6 | 0.5 | 160 \| 270±130 | CT | -5.28 | [3] |
| NWA 16001 | C3-ung | CT3 | 24 | 70 \| ~50 | sparse | none | 0.2 | 250 ± 160 | CT | -4.44 | [5] |
| Dhofar 2046 | CM2-an | CY | ~20 | ~80 | none | | | 240 ± 130 | CM (ext) | -2.58 | [4] |
| NWA 12563 | CM2-an | | 15 \| 14 | 85 \| 86 | rare (alt) | alt | | 360 | CM (ext) | -1.33 | [4] |
| NWA 14674 | CR2-an | | 20 | 80 | rare | 5 | sparse | 1300±1000 | CR (ext) | -0.55 | [4] |
| | | | | | | | | | | | |
| NWA 16476 | CO3-CO2 | CO-an | 53 \| 47-58 | 37 \| 39-49 | <4 | 10 | none/TCI | ~200 \| 190 | CO | -4.88 | [5] |
| NWA 16000 | C3-ung | CM-an | 15 | 80 | <1 | rare | 7.5 | 390±240 | CM (ext) | -1.00 | [4] |
| Ankazomena | CVox3 | | ~50 | 46 | | rare | 2.6 | 830±340 | n/a | n/a | |
| | | | | | | | | | | | |
| NWA 11085 | CO3.0 | | 51 \| 46 | 38 | 1 | none | 15 | 270 | n/a | n/a | |
| NWA 10916 | CO3.0 | | **~70** | **~30** | 1 | sparse | rare | ~200 | n/a | n/a | |
| NWA 12651 | CM2 | | ~25 | ~75 | <1 | none | PCP | ~300 | n/a | n/a | |

**Table 1: Summary of petrography of anomalous and grouped carbonaceous chondrites in this study.** The table includes the official classification of the chondrites from Meteoritical Bulletin and their potential designations based on oxygen isotope signatures. For detailed petrological descriptions, see the supplementary text. Abundances are in volume percentages, Ch = chondrule, mx = matrix, RI = refractory inclusions. The grey numbers are from Meteoritical Bulletin and the black are from this study. n/a = not analyzed, alt = altered and ext = extended. Oxygen isotope data are from 1: Metzler et al.[73], 2: R. N. Clayton and T. K. Mayeda (University of Chicago), 3: K. Ziegler (UNM), 4: C. Sonzogni and J. Gattacceca (CEREGE), 5: D. Ibarra (Brown University).

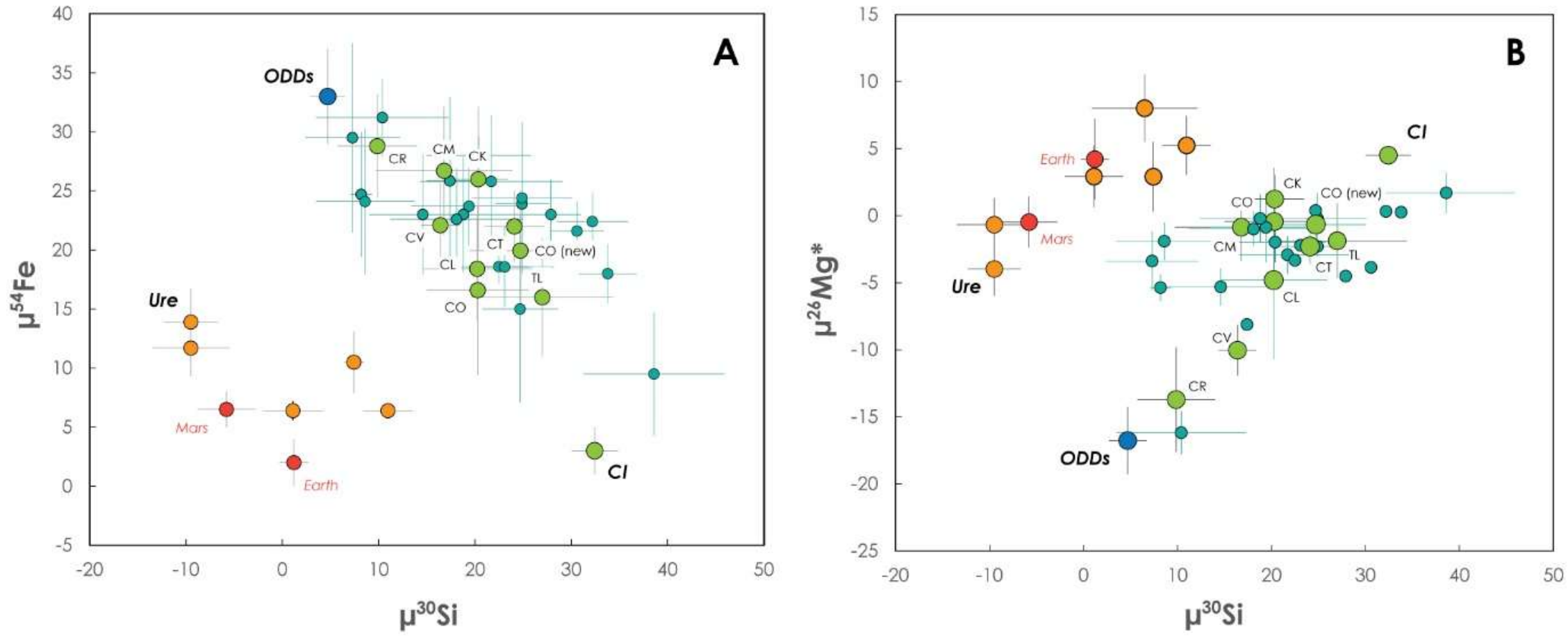

**Fig. 1: Nucleosynthetic Si, Fe and Mg isotope data of anomalous carbonaceous chondrites.** Grouped carbonaceous chondrites are light green and anomalous ones are dark green. Non-carbonaceous (a)chondrites are orange. The differentiated planetary bodies Mars and Earth are highlighted in red. Their $\mu^{54}$Fe signature reflects that of bulk silicate Earth and Mars and is affected by decoupling of Fe and Si isotope systematics due to core formation during accretion[46], explaining why these bodies plot below the CI-Ureilite correlation line. Given the contrasting geochemical behaviors of Si and Fe during core formation, their isotope signatures in BSE will decouple if accretion persists, where BSE's iron isotope composition will be skewed towards that of late accreting material. Outer disk dark clasts (ODDs)[8] are blue and weighted means of new grouplets CL, CT and a new average for CO chondrites (previous value was based on a single chondrite) are also shown as light greens spheres. **A**) $\mu^{30}$Si versus $\mu^{54}$Fe data, including data from literature [2,46]. **B**) $\mu^{30}$Si versus $\mu^{26}$Mg* data, including literature data[16,17] where $\mu^{26}$Mg* reflects values corrected to a solar Al/Mg ratio (see ***Supplementary Fig. 4-5***). Note that for Earth, Mars, Vestan meteorites and ureilites an Al/Mg ratio similar to non-carbonaceous chondrites is assumed. All error bars represent 2SE deviations from the mean.

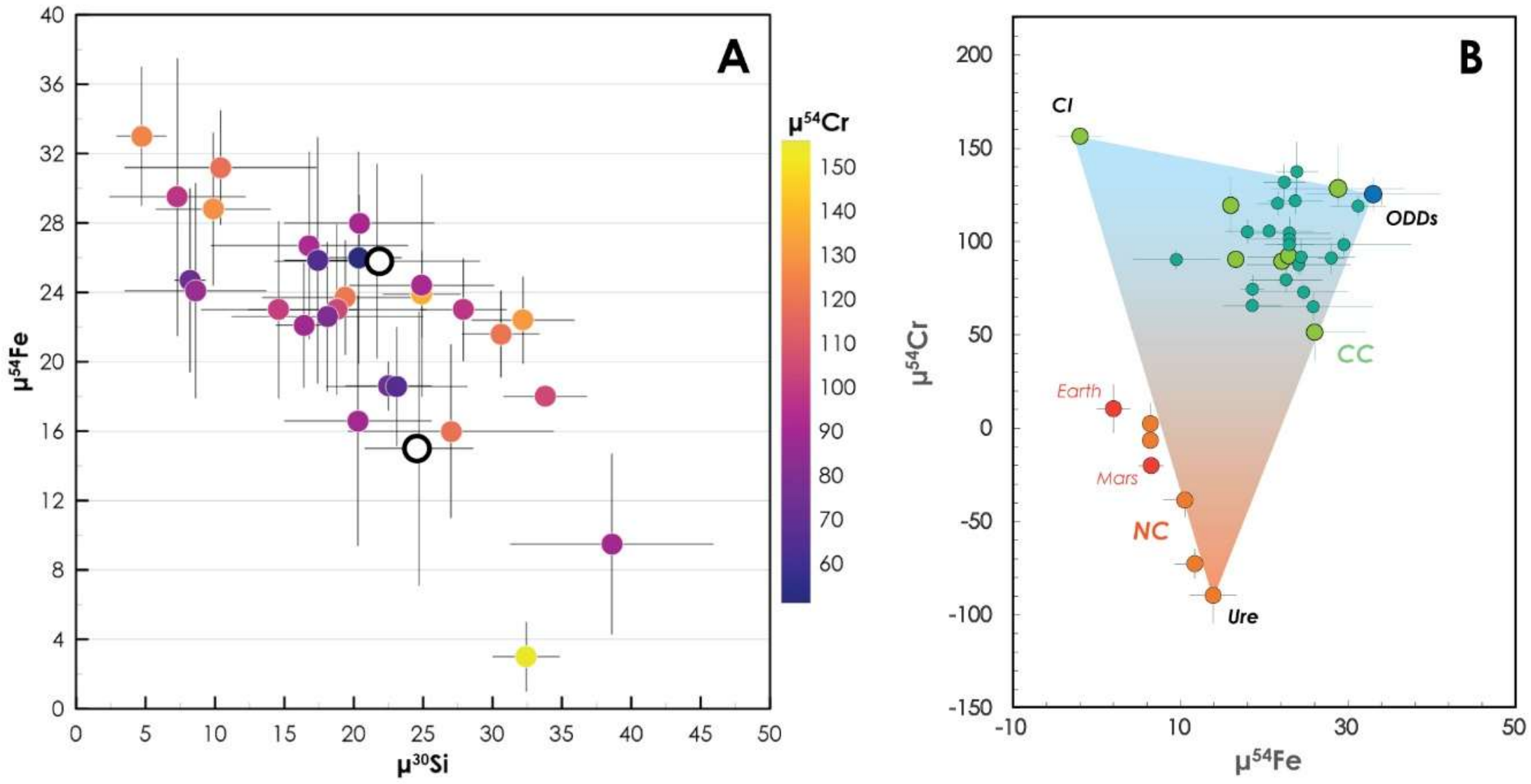

**Fig. 2: The outer disk isotope mixing line does not correlate with $\mu^{54}$Cr values. A**) The $\mu^{30}$Si versus $\mu^{54}$Fe signatures of (un)grouped carbonaceous chondrites (from ***Fig. 1A***) with color coded $\mu^{54}$Cr values for each chondrite. The open black spheres indicate 'no data' available. **B**) The $\mu^{54}$Fe versus $\mu^{54}$Cr signatures of non-carbonaceous chondrites (NC: orange spheres), grouped (light green) and anomalous carbonaceous chondrites (CC: dark green), with Cr isotope data from literature included[30,74]. The outer disk dark clasts (ODDs, blue sphere), CI chondrites and Ureilites (Ure) form a trichotomy of endmembers, like Si, Fe and Mg isotope trends in Figure 1. Notably, the $\mu^{54}$Cr signatures of CC are depleted relative to the CI-ODDs mixing line, indicating a different carrier phase for $^{54}$Cr that is more readily destroyed relative to $^{54}$Fe, $^{30}$Si and $^{26}$Mg*. Color coding in the triangle reflects increasing thermal processing from blue to red. All error bars represent 2SE deviations from the mean.

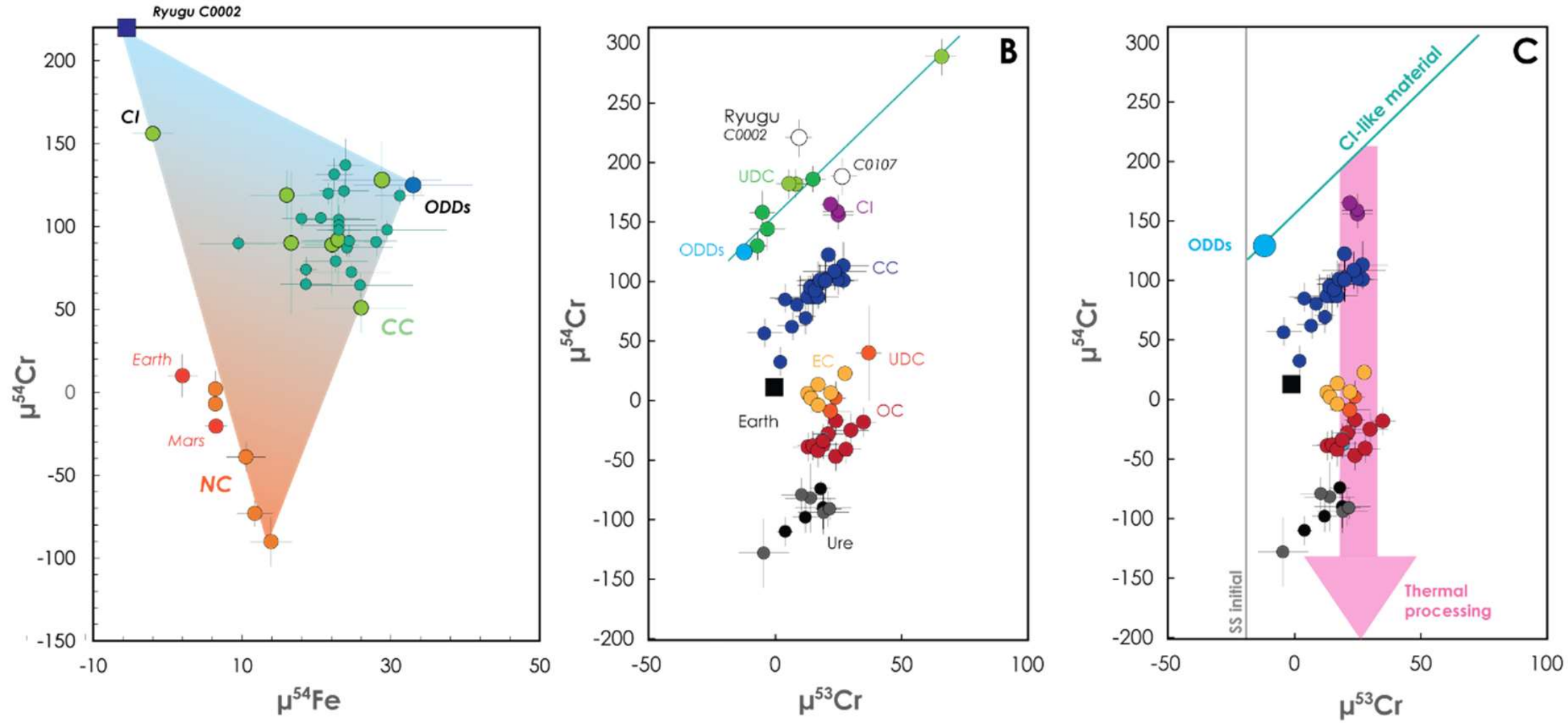

**Fig. 3: CI chondrites are thermally processed materials**. **A**) The $\mu^{54}Fe$ versus $\mu^{54}Cr$ signatures of Solar System materials (as in ***Fig. 2B***), with $^{54}Cr$-rich Ryugu particles[37] as endmember composition instead of CI chondrites. **B**) The $\mu^{53}Cr$ versus $\mu^{54}Cr$ signatures of Solar System materials, including data from literature[30,38,40,74]. Different isotope reservoirs can be observed that lie on distinct linear trends that are increasingly $^{54}Cr$-rich, including Ureilites (Ure), ordinary chondrites (OC), enstatite chondrites (EC), carbonaceous chondrites (CC) and CI chondrites. Note that outer disk dark clasts (ODDs) and CI-like ureilitic dark clasts (UDC: green spheres) plot on the most $^{54}Cr$-rich trend, together with $^{54}Cr$-rich particles from Ryugu [37]. Other UDC (orange spheres) are C2-types that contain chondrules and have a petrography related to R and OC. **C**) Assuming the ODDs-UDC line reflects the most pristine, thermally unprocessed material, CI chondrites, being depleted in $^{54}Cr$ relative to this line, possibly reflect thermally processed materials, although still more pristine than other Solar System materials. All error bars represent 2SE deviations from the mean.

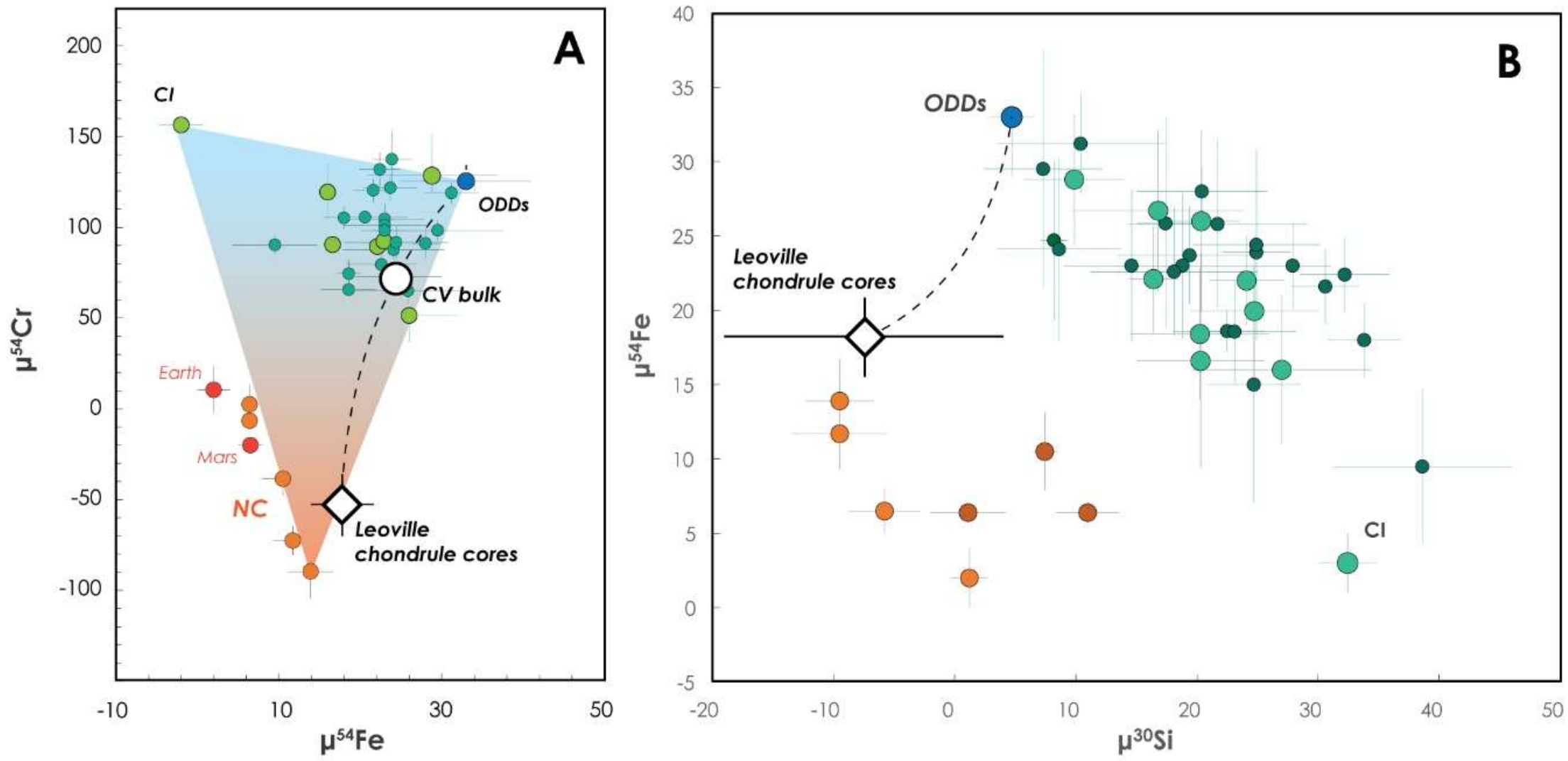

**Fig. 4: Mass transport from inner to outer disk. A**) $\mu^{54}$Fe versus $\mu^{54}$Cr signatures of Leoville chondrule cores. The $\mu^{54}$Cr values are from van Kooten et al.[43] and $\mu^{54}$Fe values from this study (see ***Supplementary Table 2***). The composition of the Leoville matrix[43] and bulk[74] is only analyzed for Cr isotopes, and the bulk CV $\mu^{54}$Fe value represents an average from literature [46] . Using mass balance, this implies a cometary nature of the matrix. Note that this mass balance is also dependent on the composition of the chondrule igneous rims, which have $\mu^{54}$Cr values intermediate to the cores and the bulk value[43]. **B**) Leoville chondrule cores in $\mu^{30}$Si versus $\mu^{54}$Fe space with $\mu^{30}$Si value from Onyett et al.[45]. The bulk $CV_{red}$ value should plot along the dashed line, although there may be an offset to the right from either igneous rims or refractory inclusions. All error bars represent 2SE deviations from the mean.

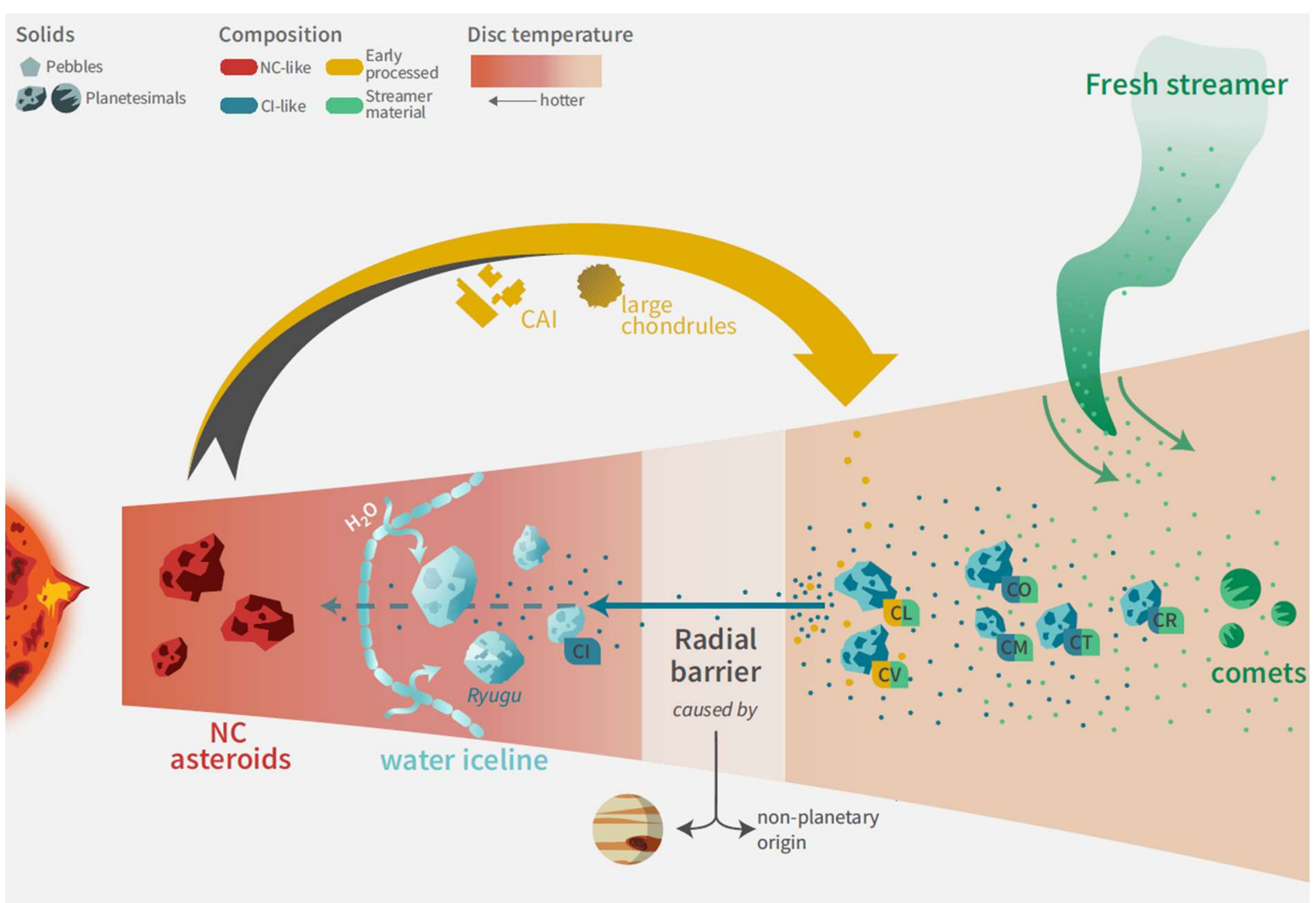

**Fig. 5: A schematic representation of the evolution of the outer disk.** CI chondrites including Ryugu and Bennu's parent bodies accrete at the water ice line inward from a barrier trapping outward transported calcium-aluminum-inclusions (CAIs) and chondrules. Fresh input from a streamer pollutes the outer disk, where mixing of CI-like and this material leads to the formation of carbonaceous chondrites.

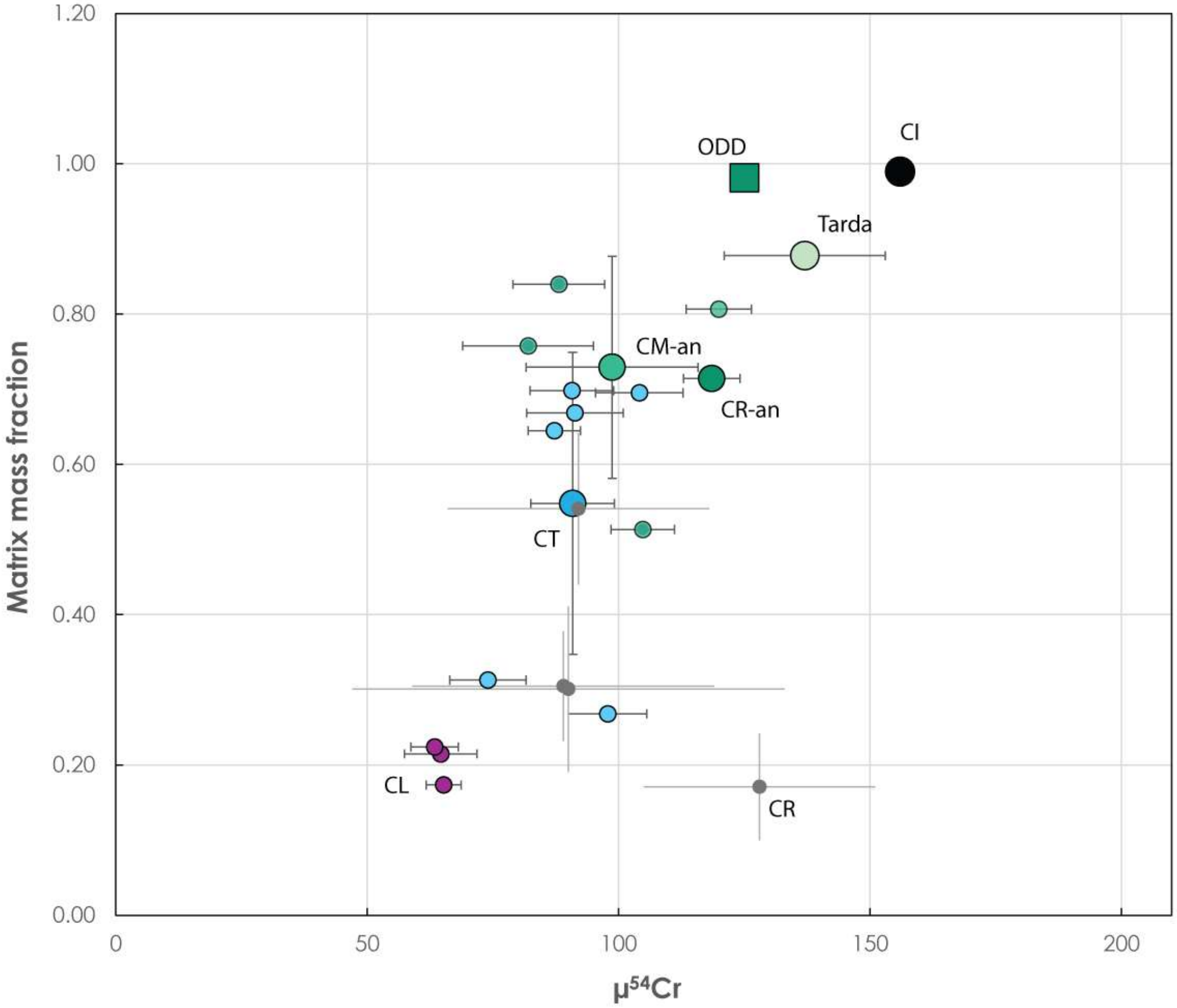

**Extended data figure 1: The $\mu^{54}$Cr compositions of chondrites in this study and from literature against their matrix mass fractions**. The average CM-an and CT chondrite composition (large spheres) is plotted with 2SE error bars from the individually analyzed chondrites (small spheres). Literature (small grey spheres) data of average CR, CM, CO, CV chondrites are also plotted, with Cr isotope data from Zhu et al.[74] and matrix mass fractions from Patzer et al.[75–77], summarized in Zhang and Grewal[78]. Matrix mass fractions from chondrites in this study are calculated from chondrule and matrix modal abundances in Table 1 and converted to mass fractions assuming a density of 3 g/cm$^3$ for chondrules and 2.5 and 3 g/cm$^3$ for type 2 and type 3 chondrites, respectively. Based on this assumption and the error in estimating chondrule and matrix modal abundances, note that the propagated error on the matrix mass fraction is estimated to be significant. The observed array with increasing $\mu^{54}$Cr values with increasing matrix mass fraction is interpreted as mixing of ODD and CI precursor material and subsequent lowering of $\mu^{54}$Cr values with increasing thermal processing during chondrule formation. Note that the CR chondrites from literature are somewhat offset from this array, similar to observations from Hellmann et al.[6], in contrast to the CR-an NWA 14674 measured in this study. This offset can be produced either by: 1) a potential mismatch between CR samples measured for $\mu^{54}$Cr and samples analyzed for their matrix mass fraction or 2) a different sequence of precursor mixing and thermal processing for CR chondrites, where first predominantly CI/Ryugu* like material is thermally processed (step 1 chondrule formation) and admixing with ODD-like material.

## Table of contents:

**Supplementary text**

## Supplementary text

### 1. Petrological descriptions

High resolution versions of sample images (***Fig. 2***) in this study are available as supplementary zip-files. Oxygen isotope data is plotted in ***Figure 1*** and is taken from Meteoritical Bulletin (see also Table 1 main text).

*1.1 CL (Loongana-type) chondrites (**Fig. 2A**):*

**NWA 13400 (CL3.9):** Large mm-sized chondrules, low-contrast between chondrules and matrix, chondrules that contain metal are heavily oxidized and many chondrules contain Fe-rich alteration rims. Many chondrules appear flattened. Fe-rich veins cut through the entire section. The matrix is entirely recrystallized and contains anhydrous silicates and sulfides or metal. Refractory inclusions were not identified in this section. Due to the low contrast, chondrule and matrix abundances and sizes were taken from the Meteoritical Bulletin.

**NWA 033 (CL4):** Wide range chondrule size distribution, with many chondrules >1 mm. High chondrule to matrix ratio, with matrix consisting of thin Fe-rich fine-grained rims around the chondrules. Chondrules are typically highly spherical, and many contain metal/sulfide rims that have been altered. The entire section is cross-cut by Fe-rich veins. Refractory inclusions were not identified in this section. Chondrule and matrix abundances and sizes were taken from the Meteoritical Bulletin.

*1.2 CT (Telekoast 001-type) chondrites (**Fig. 2B**)*

The CT group commonly has a chondrule to matrix ratio like CM chondrites but is characterized by the absence of phyllosilicates and TCIs and is defined by a specific trend in oxygen isotope space, offset from the CM chondrite line (***Fig. 1A***). CAIs are rare in these chondrites.

**NWA 16569 (C3-ung):** Closely packed small (~300 μm sized) chondrules with a relatively tight size distribution, with a few exceptions that are mm-sized. Many chondrule textures (type I and II, barred, CC, porphyritic, metal-rich). Chondrules with their pristine fine-grained rims resemble CM chondrite textures (especially Maribo without tochilinite-cronstedtite-intergrowths). Some weathering of metal grains is observed, but otherwise a low degree of alteration. Sulfides in the intra-chondrule matrix are abundant. Refractory inclusions also have fine-grained rims. Chondrule-matrix ratio was analyzed using ImageJ (***Fig. 7***).

**NWA 16559 (C3-ung):** Well-defined chondrules in a fine-grained matrix. Chondrules have a wide size distribution and contain thick fine-grained rims, although this have a low contrast with the intra-chondrule matrix. No clear signs of alteration are observed in the matrix and it has low abundances of (large) sulfides. Metal or sulfide are present in the larger chondrules. Many of the larger chondrules are compound chondrules and have irregular shapes. Refractory inclusions were not identified in this section. Chondrule-matrix ratio was analyzed using ImageJ (***Fig. 7***).

**Chwichiya 002 (C3.00-ung):** This chondrite appears very heterogeneous and brecciated, with abundant chondrule fragments. Chondrules vary in size (typically <500 μm) and appearance (with or without fine-grained rims, irregular and spherically shaped, type I and II, some metal-containing, etc.) and the matrix shows light zoning of sulfide-rich and poor areas on a mm-scale. Chondrule and matrix abundances and sizes were taken from the Meteoritical Bulletin.

**NWA 14051 (C3-ung):** Similar texture to Chwichiya 002, with a tendency towards larger chondrule sizes (one chondrule >1 mm). Local Fe-rich regions indicate alteration zones where metal is weathered/altered. The matrix is very porous and contains abudant small cracks and individual tiny metal/sulfide grains are observed. Chondrule and matrix abundances and sizes were taken from the Meteoritical Bulletin.

**NWA 14139 (C3-ung):** Well-defined chondrules in a fine-grained matrix. Chondrules have a wide size distribution (typically <500 µm) and many contain fine-grained rims. Metal grains in the matrix show limited indication of alteration. Most of the matrix is very fine-grained with sulfides interspersed with silicates. Chondrules present as relatively irregular compound textures. Fragmented chondrules are observed throughout the section and many are type II fragments. Chondrule-matrix ratio was analyzed using ImageJ (***Fig. 7***).

**NWA 16001 (C3-ung):** Brecciated chondrite (different degrees of alteration: Fe-rich and poor zones) on the cm-scale, texture is otherwise similar to NWA 14139. Well-defined chondrules and sparse CAIs in a fine-grained matrix. We could identify some fine-grained matrix lumps. Chondrule-matrix ratio was analyzed using ImageJ (***Fig. 7***).

*1.3 CM-anomalous chondrites (**Fig. 2C**)*

The CM-anomalous chondrites are defined here by their low chondrule-matrix ratios and small chondrules sizes (~300 µm). These chondrites do not exhibit typical TCIs as do classical CM chondrites. Their oxygen isotope signatures lie on a more $^{16}$O-poor extension of the CM trend (***Fig. 1B***).

**NWA 16554 (C3-ung):** Two separate handpieces of this meteorite were sampled. Small chondrules (<300 µm) in low contrast to the matrix, the latter being highly porous, resulting in cracks that surround each chondrule. The chondrules themselves also contain voids. Chondritic texture resembles the anomalous chondrite Bells. Sulfide and metal grains are present throughout the section and are also included in chondrules. Initial classification suggests this sample to represent a thermally metamorphosed and dehydrated C2 chondrite (Meteoritical Bulletin). Chondrule-matrix ratio was analyzed using ImageJ (***Fig. 7***).

**Dhofar 2046 (CM2-an):** Well-defined small chondrules (<500 µm) in a fine-grained matrix. Approximately half of the chondrule volume is taken up by a few larger ~500 µm sized chondrules, whereas the remaining could be defined as microchondrules or fragments. Chondrules are typically surrounded by a dark fine-grained rim relative to the lighter matrix. The matrix is a mixture of dark patches resembling the FGRs and lighter patches that are more sulfide-rich. The total matrix volume is relatively large (80%) and determined using ImageJ. This chondrite has been suggested as a potential CY chondrite[1,2].

**NWA 12563 (C2-ung1):** The petrography of this chondrite has been described in detail by Hewins et al.[3], who characterized this chondrite as an anomalous CM chondrite with a similar chondrite to matrix ratio and chondrule size distribution but TCIs being absent from the matrix. The meteorite has been linked to Bells and Essebi, and potentially to material returned from the asteroid Bennu. From Hewins et al.[3]: The chondrule mesostasis has been replaced by serpentine-saponite and chlorite, and the metal by serpentinite in the type I chondrules. We observe the replacement of mesostasis and metal as well but have not identified the mineralogy of the secondary grains. Chondrule-matrix ratio determination using ImageJ agrees with the ratio found by Hewins et al.[3].

**NWA 16000 (C3-ung):** Well-defined chondrules in an abundant fine-grained matrix (80 vol.%). Chondrules have a wide size distribution (some >1 mm) and contain thick fine-grained rims, most of which have a high contrast with the sulfide-rich matrix. Sulfides are relatively large and clearly discernible in the intra-chondrule matrix, which is heterogeneous with darker patches (lower sulfide content) and lighter patches (higher sulfide content). Chondrule fragments (both type I and type II) are abundant. Chondrule-matrix ratio was analyzed using ImageJ.

*1.4 CO and CV chondrites (**Fig. 2D**)*

**NWA 11085 (CO3.0):** Approximately 50-50 chondrule-matrix distribution with relatively small chondrules (<500 μm). Some chondrules contain fine-grained rims that are not easily distinguishable from the intra-chondrule matrix in texture. Larger metal grains are spread throughout the section and are all weathered to some extent. The initial classification of this chondrite in Meteoritical Bulletin describes this chondrite as strongly weathered, which is not obvious in this section. Chondrule-matrix ratio was analyzed using ImageJ.

**NWA 10916 (CO3.0):** Approximately 60-40 chondrule-matrix distribution with relatively small chondrules (<500 μm). Some chondrules contain fine-grained rims that are not easily distinguishable from the intra-chondrule matrix in texture. Larger metal grains are spread throughout the section and are all weathered/altered to some extent. The chondrite contains many chondrule fragments (type I and type II). Chondrule-matrix ratio was analyzed using ImageJ.

**NWA 16476 (CO3.0-CO2-an):** Initial characterization of this chondrite showed two lithologies: one that included TCIs and fine-grained rims around chondrules resembling CM chondrites (lithology A), and one without TCIs but including rounded metal blobs in chondrules and matrix and without FGRs resembling CO chondrites. Our section only includes lithology B related to CO chondrites. The samples looks very pristine and shows no significant signs of weathering. Chondrule-matrix ratio was analyzed using ImageJ.

**Ankazomena (Cvox3):** In the initial classification, this sample has been described as a Bali-type oxidized CV chondrite, with relatively large mm-sized chondrules (<5 mm, ~50 vol.%) set in a fine-grained, recrystallized matrix. The overall texture of this chondrite resembles Bali and Raman-derived spectra of this sample (Camplong et al., 2026) reflect thermal maturation of organic matter with a petrologic type of >CV3.6[4].

*1.5 CR-an and SAH00182 (**Fig. 2E**)*

**NWA 14674 (CR2-an):** Highly brecciated chondrite with mostly mm-sized chondrules set in abundant fine-grained matrix that contains a high (framboidal) magnetite content. High abundance of matrix (~80 vol.%) sets this chondrite apart from CR chondrites. Dark inclusions and type II chondrules are abundant. Chondrules typically contain Fe-rich fine-grained rims. The petrography of this anomalous CR2 is described in detail in Hewins et al.[5]. Chondrule-matrix ratio was analyzed using ImageJ.

**SAH 00182 (C3-ung):** Densley packed, mm-sized chondrules (up to 6 mm in size) in a metal-rich anhydrous matrix. The largest chondrules are cryptocrystalline and have single bow-shaped dents. Otherwise, the chondrite texture resembles reduced CV-type chondrites to a high degree, in agreement with the oxygen isotope composition. The exception is that CAIs are not abundant in this chondrite. Chondrule-matrix ratio was analyzed using ImageJ. For more discussion on the classification of this chondrite, the reader is referred to Metzler et al.[6].

*1.6 Tarda: a unique fall (**Fig. 2F**)*

**Tarda (C2-ung):** Tarda is a well-known unique fall, that has been described in many literature (Marrocchi et al., 2021). It has a high matrix abundance (93 vol.%) that appears homogeneous and predominantly contains phyllosilicates, and minor magnetite, sulfides, carbonates and olivine. Chondrules are typically small (<300 μm) and can contain fine-grained rims. The chondrule-matrix ratio was adapted from Marrocchi et al.[7].

## 2. The provenances of carbonaceous chondrite grouplets

We have analyzed the mass-dependent and mass-independent Si, Mg, Fe and Cr isotope compositions using multi-collector inductively coupled plasma mass spectrometry (MC-ICPMS, see Methods, ***Table 1*** and *raw data table R2*). All mass-independent isotope data is reported in the μ-notation, signifying parts per million deviations from the terrestrial standard value. Notably, the reported $\mu^{26}Mg^*$ (the decay product of $^{26}Al$) values have been corrected to their solar $^{27}Al/^{24}Mg$ ratio (see supplementary materials, ***Fig. 4-5, Table 1***), to avoid artificially high $\mu^{26}Mg^*$ signatures from calcium-aluminum-inclusion contributions[8].

For the six CT chondrites, the weighted mean (WM) of $\mu^{30}Si$ is 24.1 ± 3.1 ppm (2SE for all reported errors), with an outlier of 8.4 ± 5.1 ppm for NWA 14051. The $\mu^{54}Fe_{WM}$ of 22.0 ± 3.0 ppm (2SE, no outliers), the $\mu^{54}Cr_{WM}$ of 89.6 ± 7.8 (2SE, no outliers) and $\mu^{26}Mg^*_{WM,\ SOLAR}$ of –2.3 ± 1.3 (2SE, no outliers) tie the CT chondrites closely to the CM group[8–12]. From a petrological perspective, CM and CTs mainly differ in their matrix mineralogy (i.e., tochilinite-cronstedtite-intergrowths are absent in CT), and the latter possibly have a larger range of chondrule/matrix ratios and chondrule size distributions. Their deviating oxygen isotope signatures have promoted the view that CT and CM chondrites have distinct parent bodies, which is not confirmed by the present data.

The Si, Fe and Mg isotope signatures for the CL chondrites are unresolvable from CT and CM chondrites ($\mu^{30}Si_{WM}$ = 18.4 ± 4.4 ppm, $\mu^{54}Fe_{WM}$ = 23.6 ± 5.5 ppm, $\mu^{26}Mg^*_{WM,\ SOLAR}$ = –4.8 ± 5.9 ppm), but their $\mu^{54}Cr_{WM}$ signature is distinct (65.1 ± 3.1 ppm), in agreement with previously reported values[6]. This distinction aligns this new grouplet more closely with CV and CK chondrites, which agrees with their relatively large chondrule sizes, chondrule/matrix ratio and metal abundance, but not with their relative depletion of CAIs relative to CV chondrites.

The CM-anomalous chondrites show larger isotope variations ($\mu^{30}Si_{WM}$ = 30.2 ± 8.2 ppm, $\mu^{54}Fe_{WM}$ = 20.1 ± 6.4 ppm, $\mu^{54}Cr_{WM}$ = 90.4 ± 13.2 ppm), except for the $\mu^{26}Mg^*$ values that show a tight range ($\mu^{26}Mg^*_{WM}$ = –0.3 ± 1.7 ppm) similar to CM ($\mu^{26}Mg^*_{WM,\ SOLAR}$ = –1.1 ± 1.9 ppm)[8,9] and CO chondrites studied here ($\mu^{26}Mg^*_{WM,\ SOLAR}$ = –1.4 ± 2.3 ppm). The CM-an individual isotope signatures (***Table S1***) are resolvable from one another and appear to reflect non-uniform accretion on one or more CM-anomalous parent bodies. Their isotope composition is also distinct from Bells, with $\mu^{54}Cr$ = 137 ± 13 ppm and $\mu^{26}Mg^*_{SOLAR}$ = –9.7 ± 0.9 ppm[9], which is thought to represent a transitional accretion phase to CM and CR chondrites[9,13]. Similarly, both SAH 00182 and Ankazomena are isotopically intermediate to CM and CR chondrites. The CR-an NWA 14674, which has an unusually low chondrule/matrix ratio, has clear affinities with CR chondrites. The Tarda meteorite has not been associated with any grouplet, but has close affinities to Tagish Lake, CI and CY chondrites in oxygen isotope space[7]. In Si, Fe and Mg isotope space, Tarda is closely linked to CT chondrites, although its $\mu^{54}Cr$ signature is significantly higher (137 ± 16 ppm). However, previous studies show that Tarda is likely heterogeneous in terms of its Cr isotope signature[14–16].

The CO chondrites are enriched in $^{30}Si$ relative to CM chondrites ($\mu^{30}Si_{WM,\ CO}$ = 28 ± 4.7 ppm), but show similar $\mu^{54}Fe_{WM}$ and $\mu^{26}Mg^*_{WM,\ SOLAR}$ values ($\mu^{26}Mg^*_{WM,\ SOLAR}$ = –1.4 ± 2.3 ppm, $\mu^{54}Fe_{WM}$ = 23.3 ± 3.9 ppm). For these isotope systems, CO chondrites are most closely matched to CT chondrites.

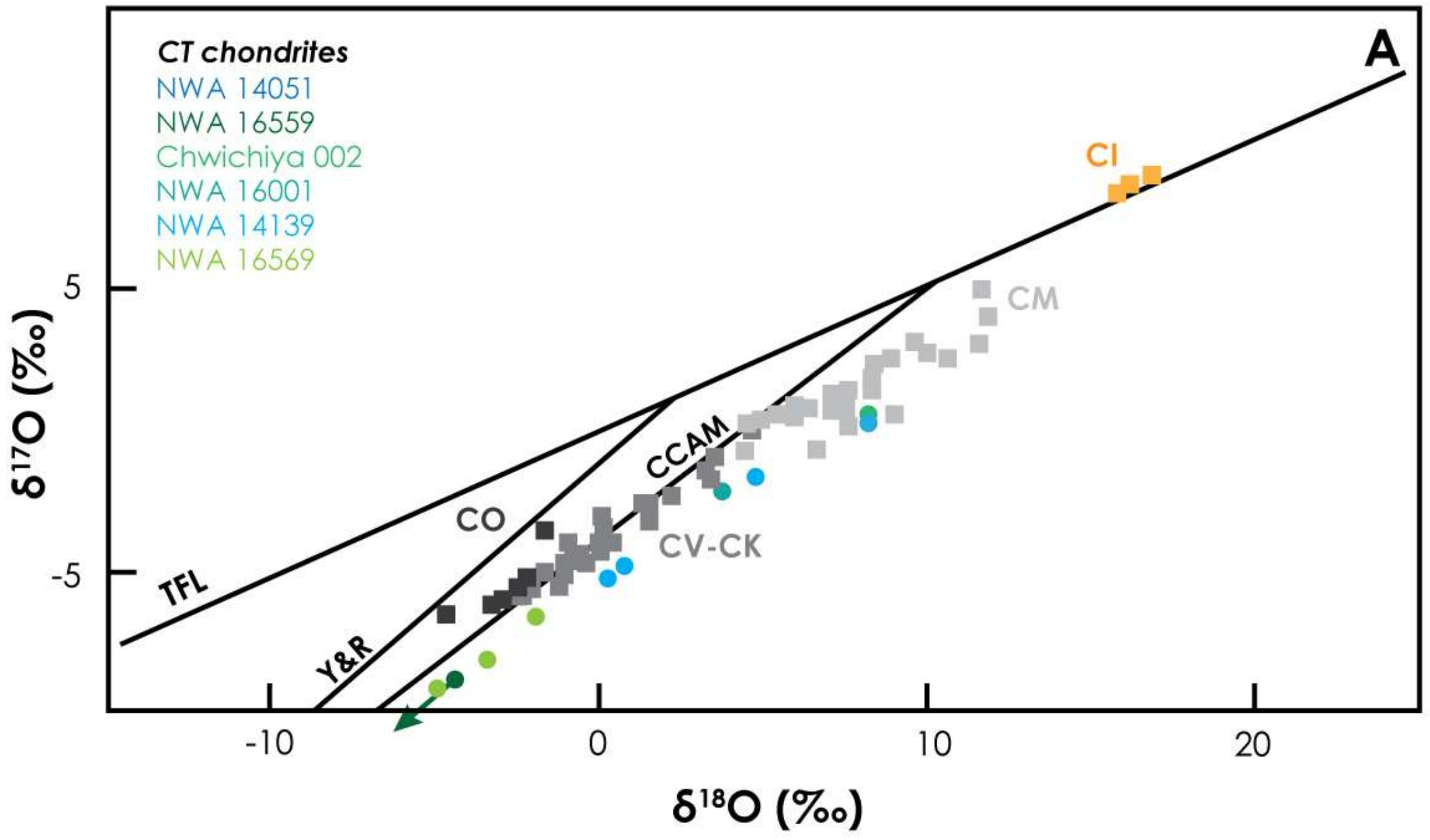

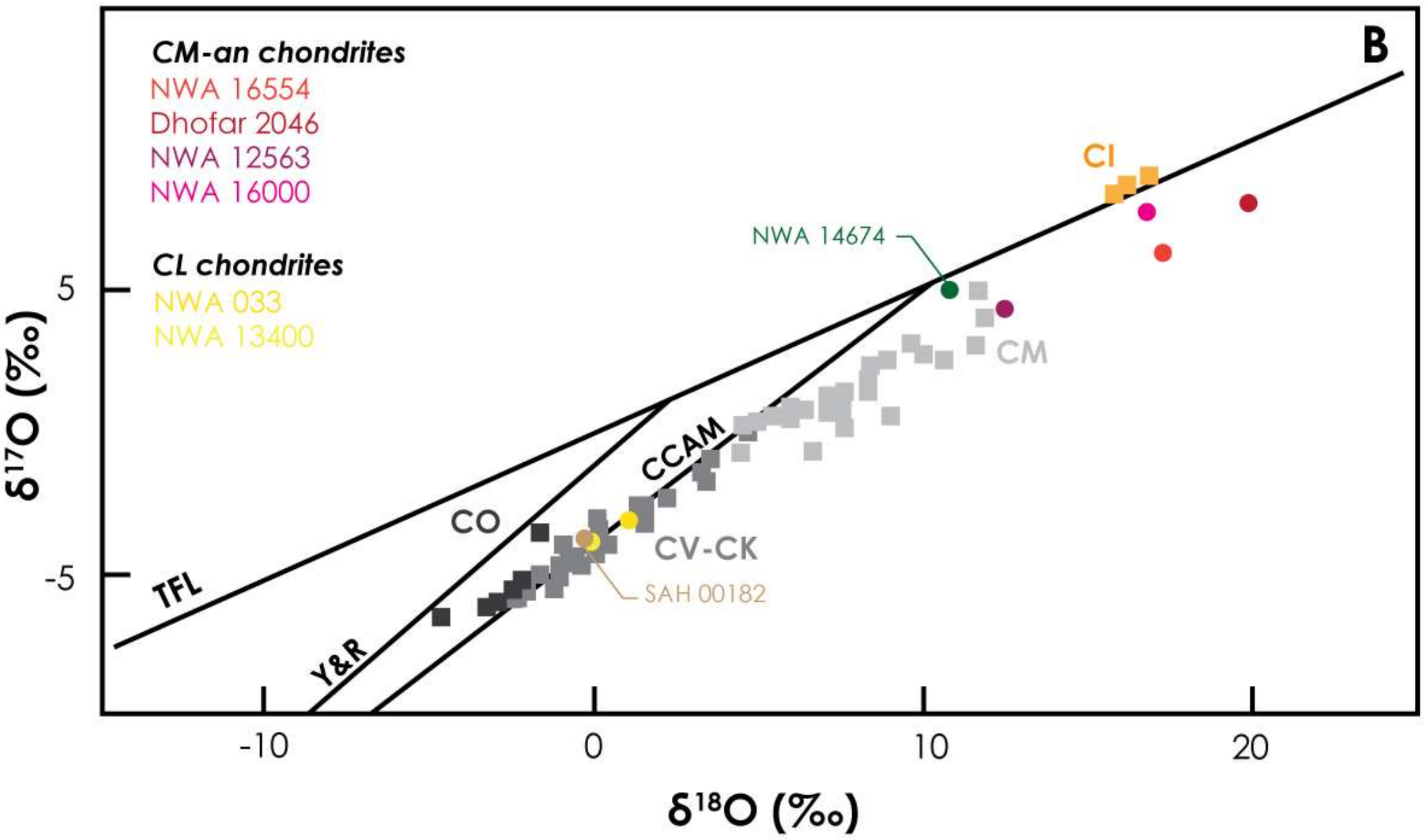

**Fig. 1: Three oxygen isotope plots** including **A**) CT chondrites from this study, **B**) CM-an and CL chondrites, SAH 00182 (C3-ung) and NWA 14674 (CR-an). For reference, we also plot the terrestrial fractionation line (TFL), the Young and Russel slope-1 line (Y&R) and the carbonaceous chondrite anhydrous mineral line (CCAM), as well as CO, CM, CI and CV-CK chondrites. The plot uses data from the Meteoritical Bulletin, including literature values[17–19].

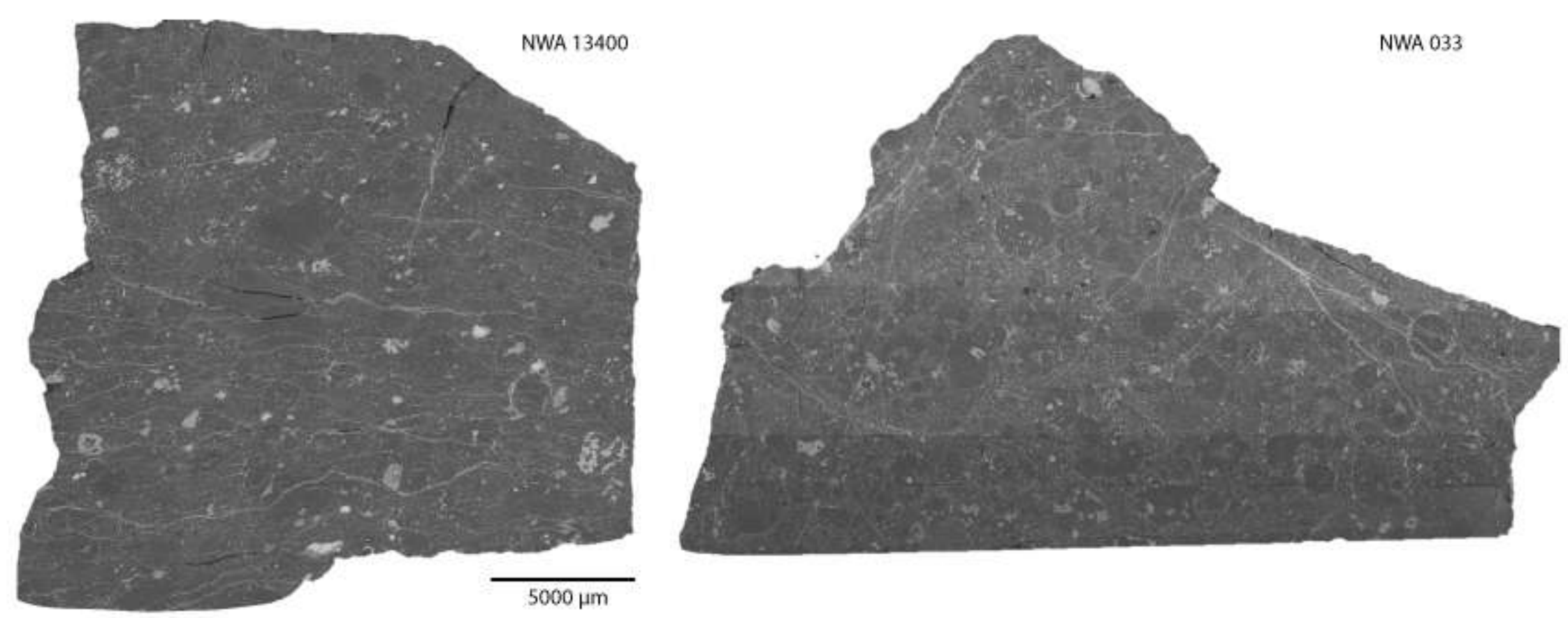

**Fig. 2A: BSE images of Loongana-type (CL) carbonaceous chondrites**

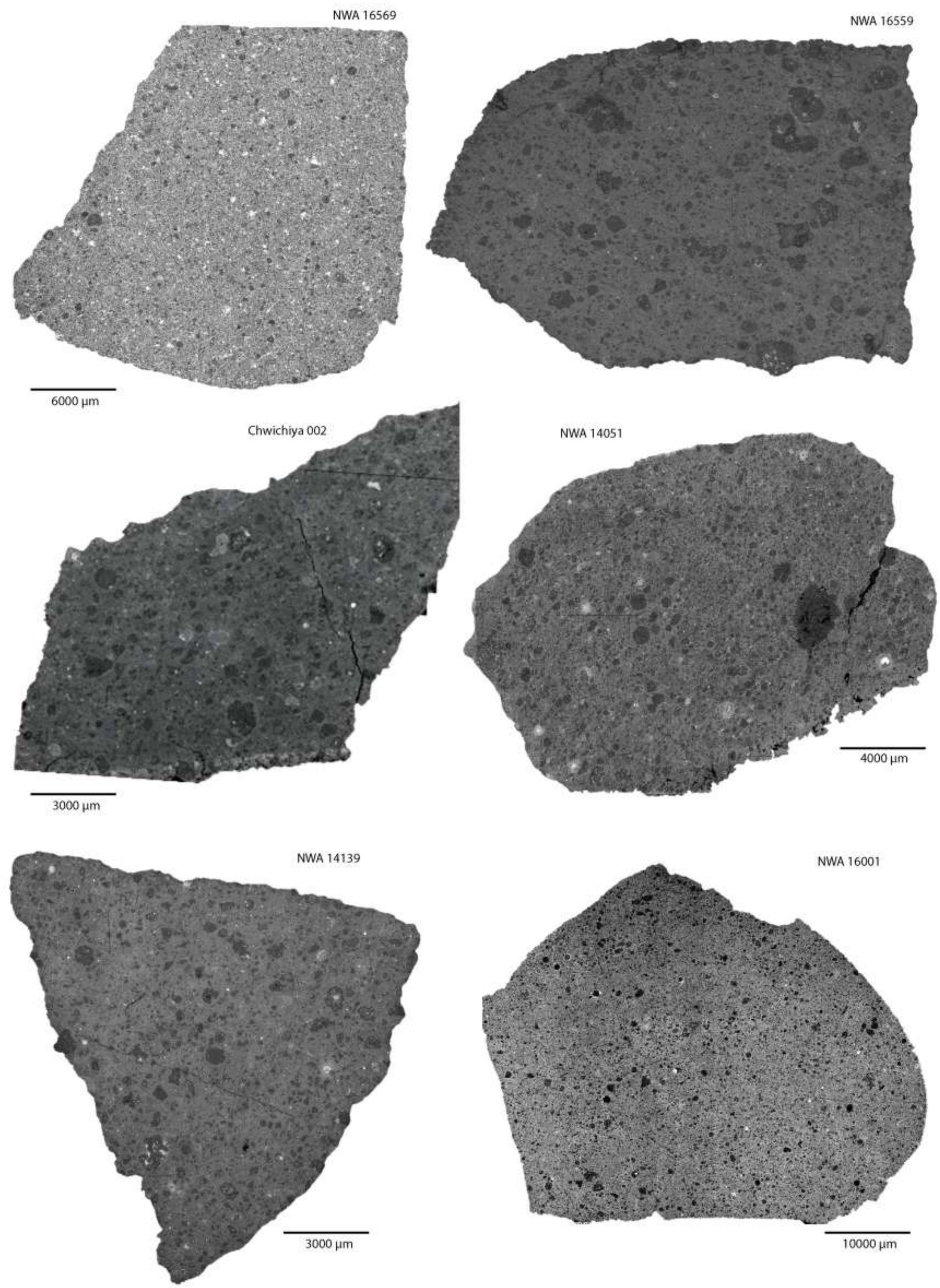

**Fig. 2B: BSE images of Telekoast 001-type (CT) ungrouped chondrites**

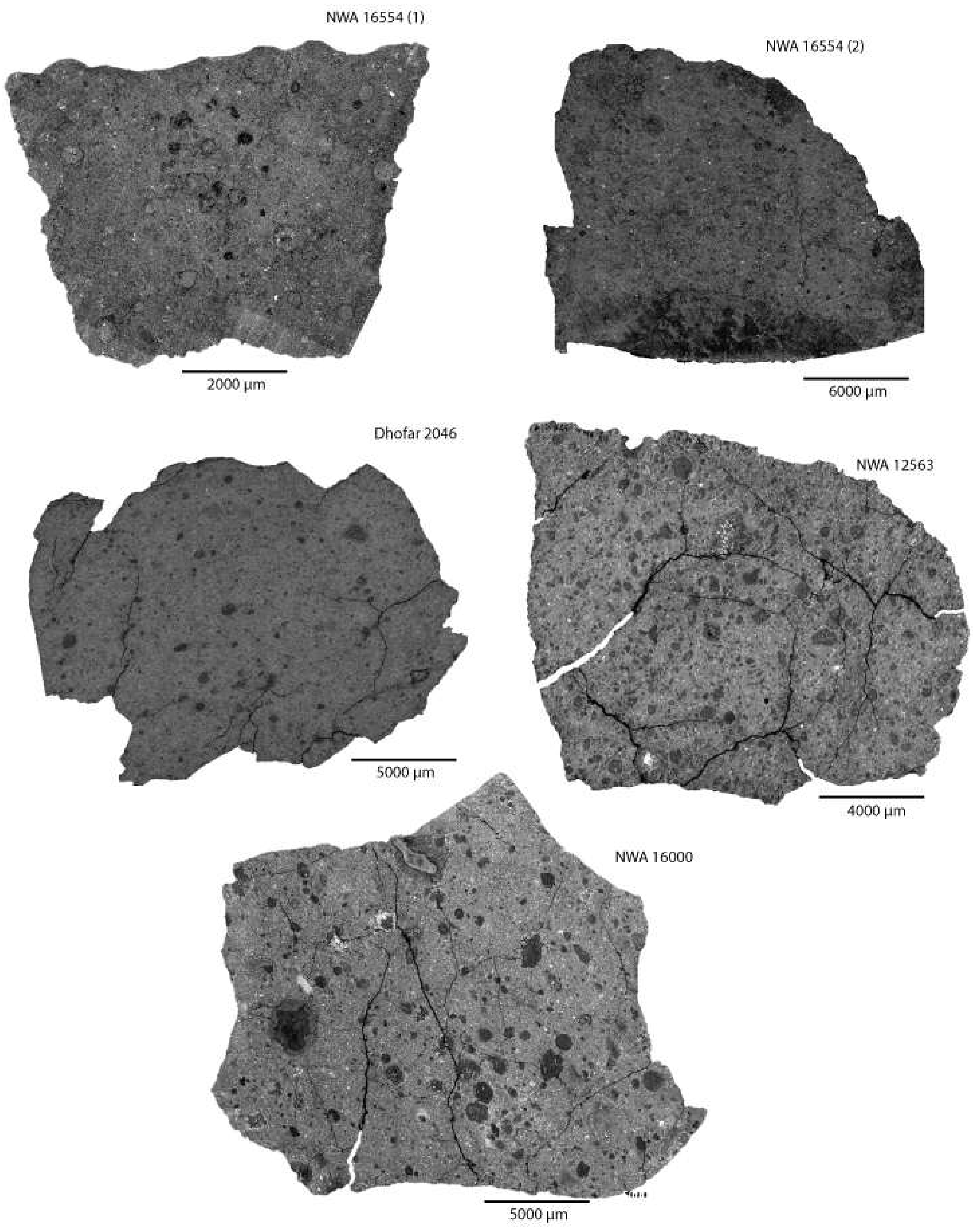

**Fig. 2C: BSE images of (potential) CM-anomalous chondrites.**

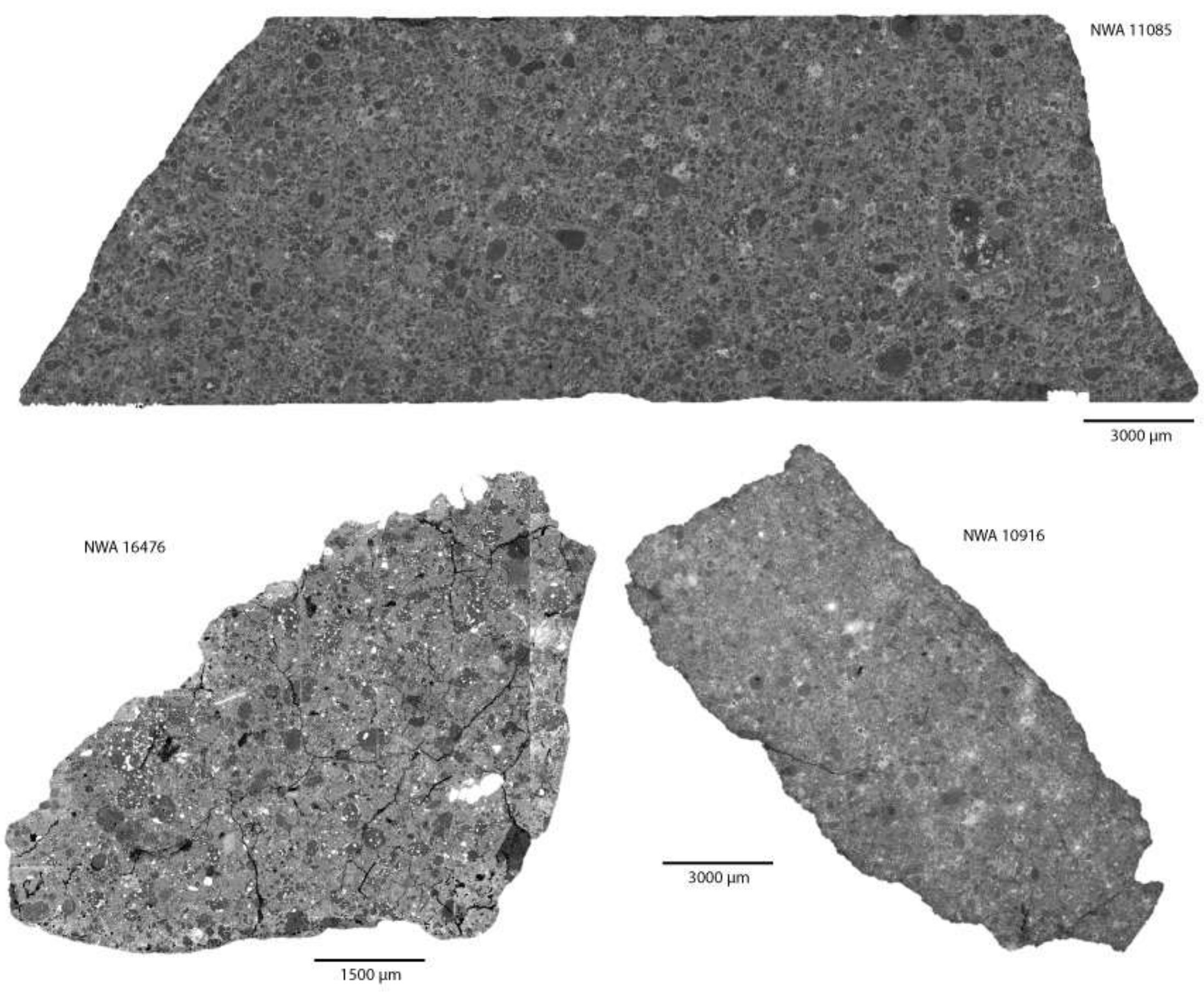

**Fig. 2D: BSE images of CO chondrites.** Note: NWA 16476 is potentially CO-anomalous.

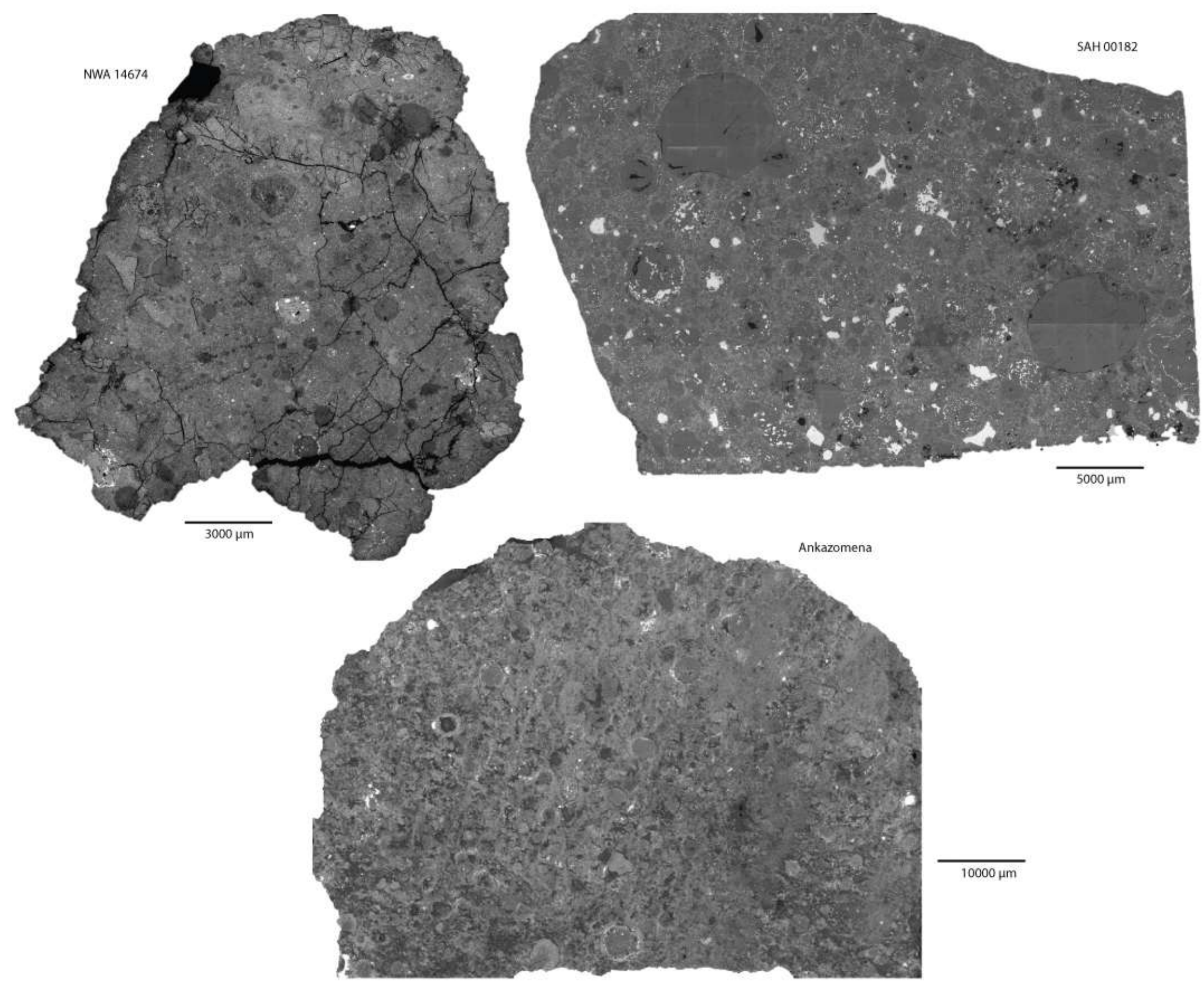

**Fig. 2E: BSE images of CR-an NWA 14674, C3-ung SAH 00182 and Ankazomena.**

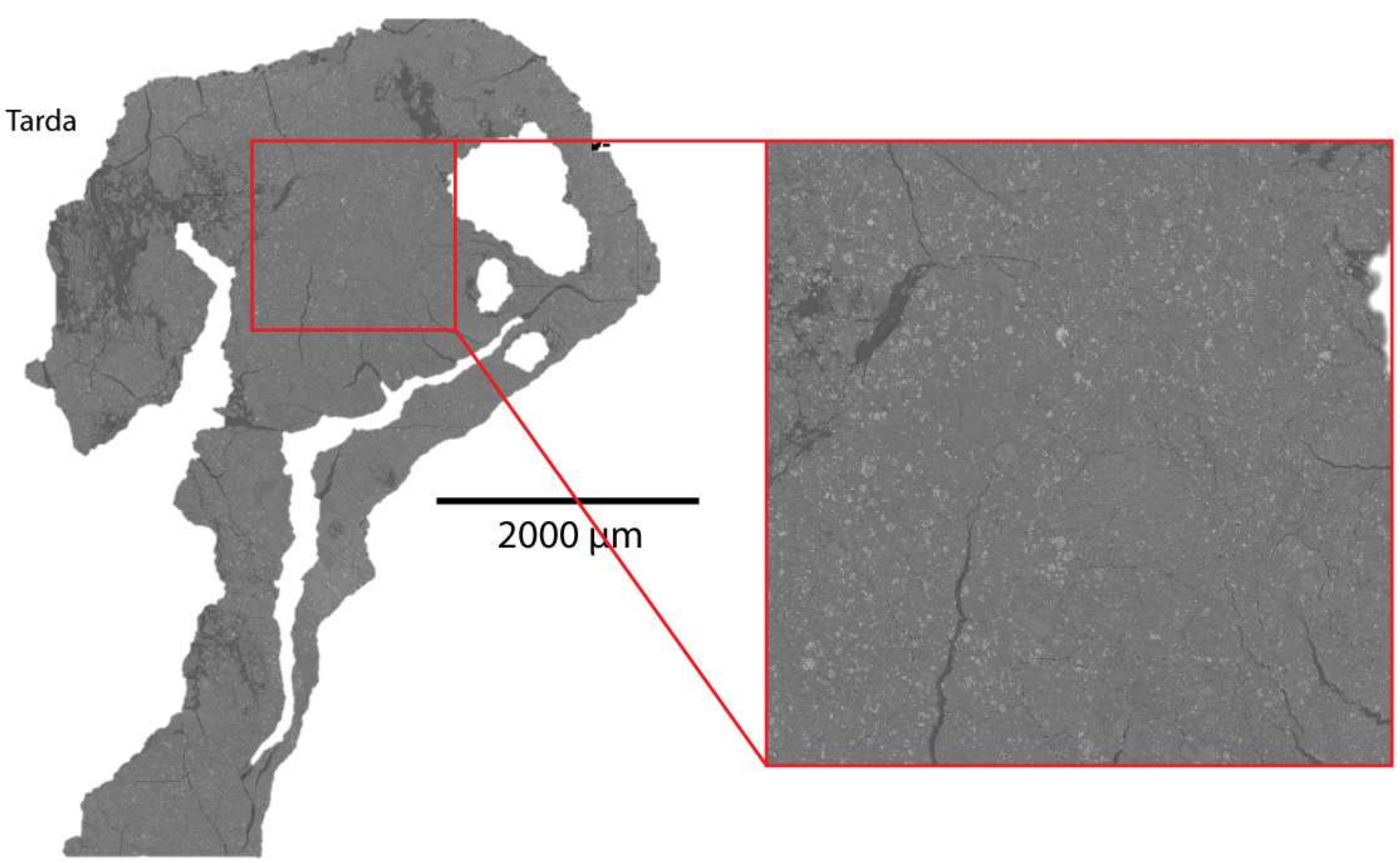

**Fig 2F: BSE images of Tarda.** Note that the sample has a high matrix content and is very fragile and thus hard to mount.

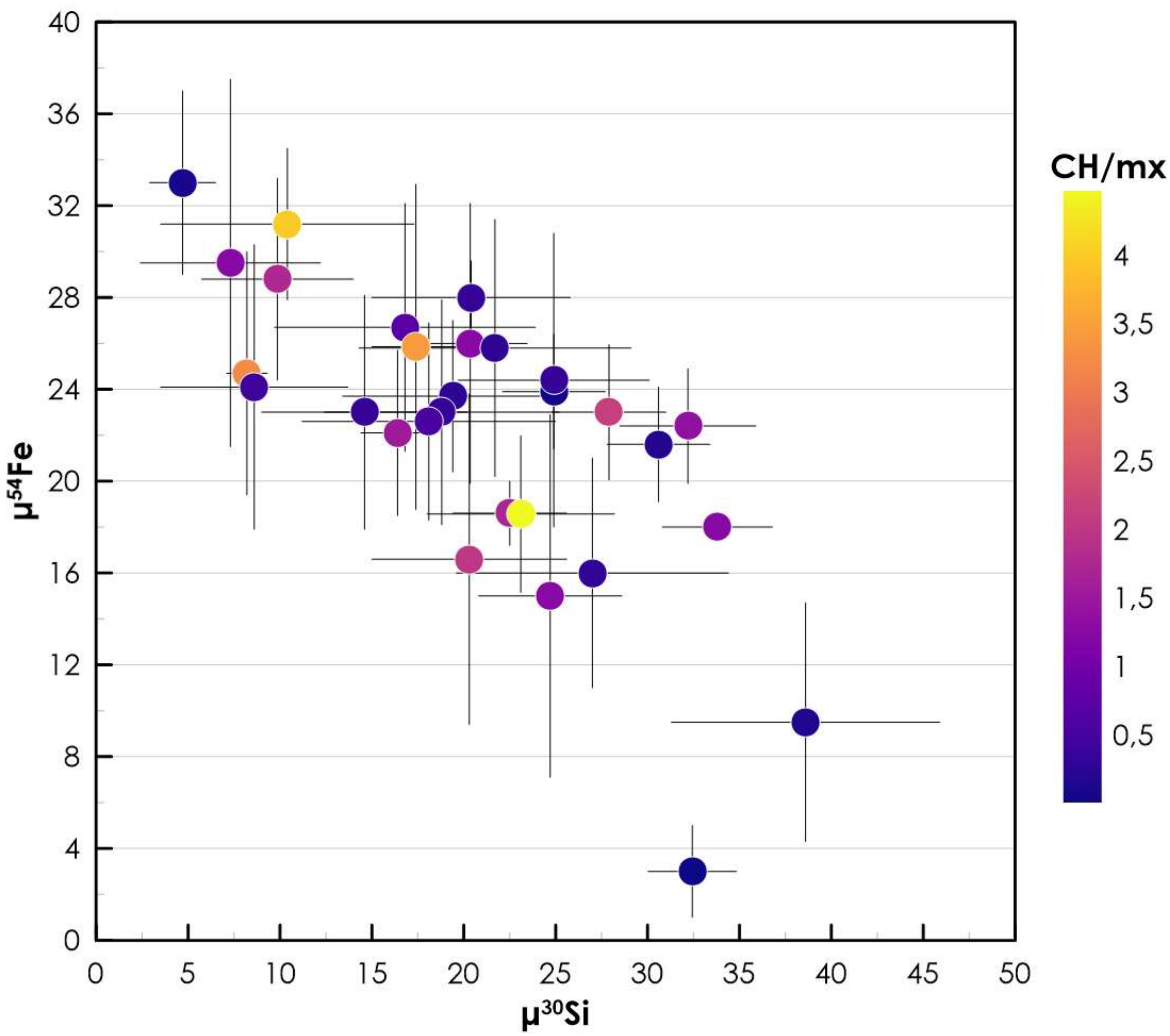

**Fig. 3: The μ³⁰Si versus μ⁵⁴Fe values of ungrouped carbonaceous chondrites.** All data is color coded according to their chondrule to matrix ratios. There is no obvious relationship between this ratio and the variations in nucleosynthetic signatures. Error bars represent 2SE deviations from the mean.

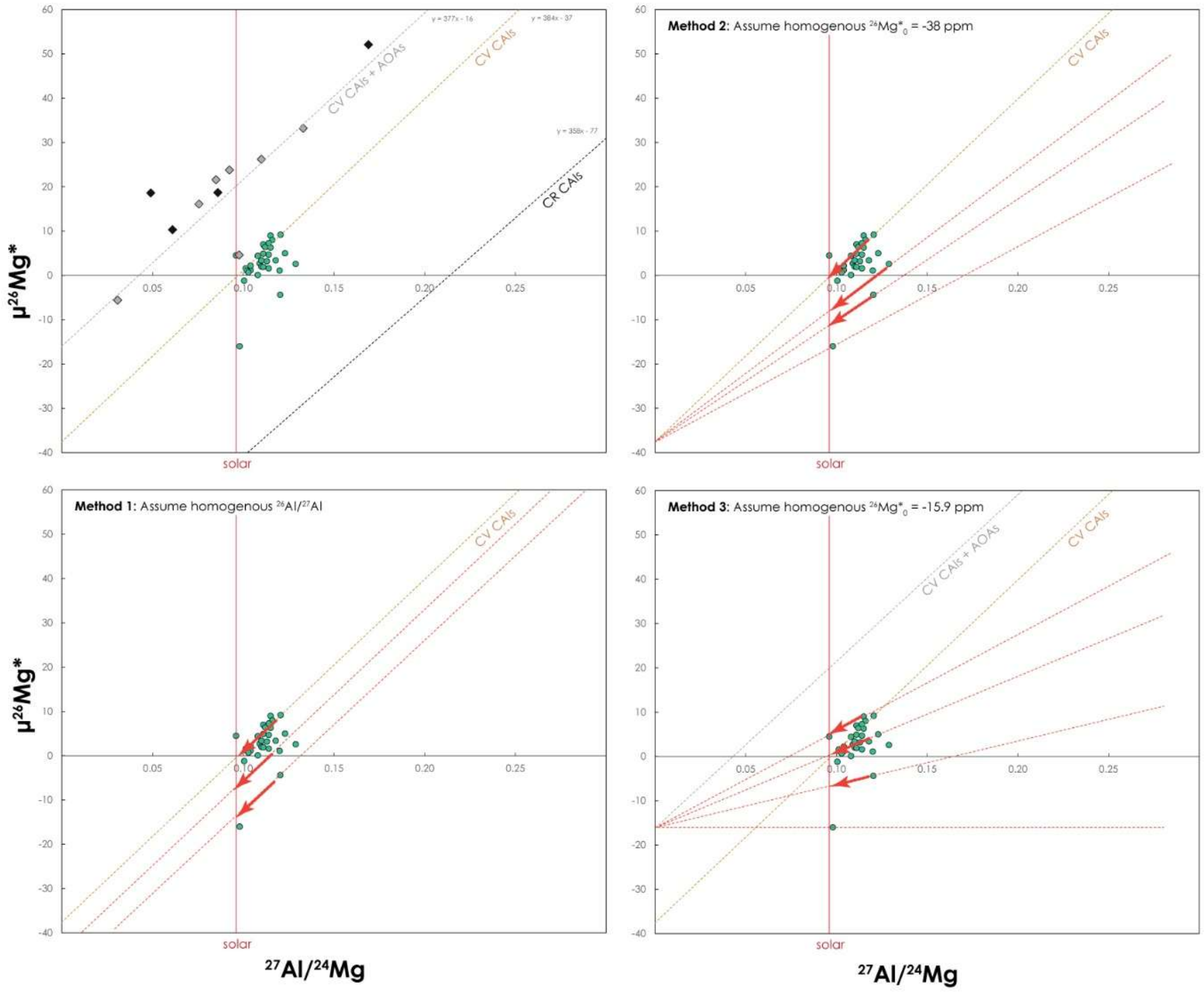

**Fig. 4: Different methods to correct Mg isotopes and compare data. A**) $^{27}Al/^{24}Mg$ ratios versus measured $\mu^{26}Mg^*$ values of ungrouped chondrites (green spheres), CV ameboid olivine aggregates (AOAs, grey diamonds), CR AOAs (black diamonds), the CV CAI + AOA regression line[20], the CV CAI regression line[20], the CR CAI regression line[21]. The solar $^{27}Al/^{24}Mg$ ratio of 0.096[22] is marked with a red dashed line. **B**) Correction method 1: we assume a homogeneous and canonical (CAI-like) initial $^{26}Al/^{27}Al$ ratio, which results in correction slopes parallel to the CV CAI line. Note that this method corrects best for the ingrowth of CAI-like materials and is used to compare data ($\mu^{26}Mg^*_{SOLAR}$) in the main text (see also ***Fig. 5A***). **C**) Correction method 2: We assume a fixed $\mu^{26}Mg^*_0$ value of -38 ppm[23] and heterogeneously distributed initial $^{26}Al/^{27}Al$ ratio, resulting in two-point slopes that are used to calculate the $\mu^{26}Mg^*_{SOLAR}$. The resulting values are nearly identical to method 1 values (***Fig. 5B***). **D**) Correction method 3: We assume a fixed $\mu^{26}Mg^*_0$ value of -15.9 ppm[20] and heterogeneously distributed initial $^{26}Al/^{27}Al$ ratio, resulting in two-point slopes that are used to calculate the $\mu^{26}Mg^*_{SOLAR}$. This correction removes a significant part of the observed $\mu^{26}Mg^*$ variability (***Fig. 5C***), suggesting that variability between NC and CC are the result of $^{26}Al$ heterogeneity. Since this method is not good at removing CAI contribution, we prefer method 1 to compare $\mu^{26}Mg^*$ values. Note that this correction assumes that Al/Mg fractionation occurred at the time of CAI formation. If this assumption is incorrect, it may result in the introduction of variability in the corrected $\mu^{26}Mg^*$ values.

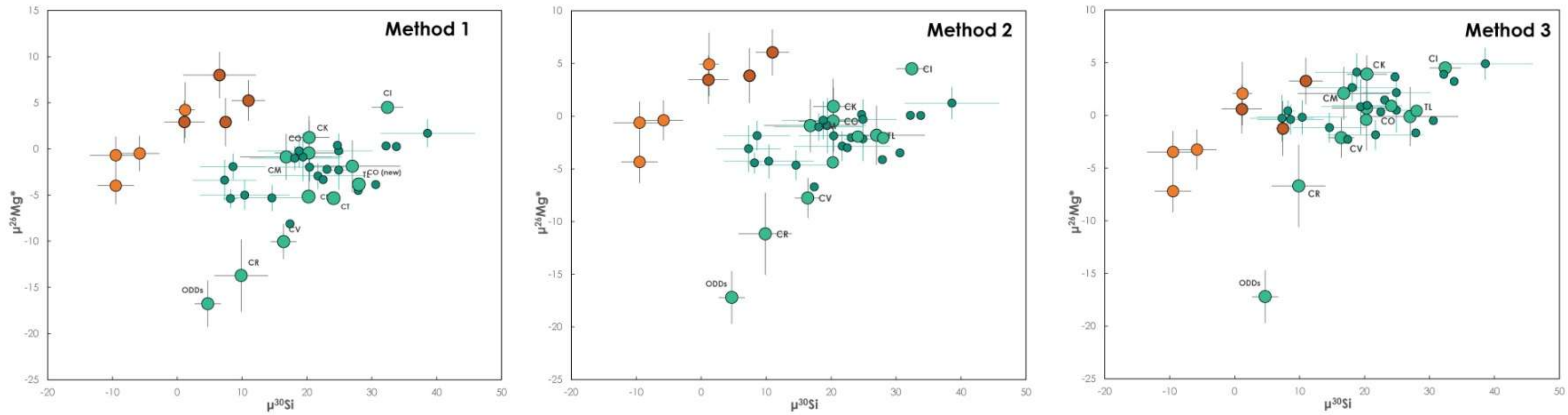

**Fig. 5: Results of three different correction methods in $\mu^{30}$Si versus $\mu^{26}$Mg* space.** Method 1 results are discussed in the main text.

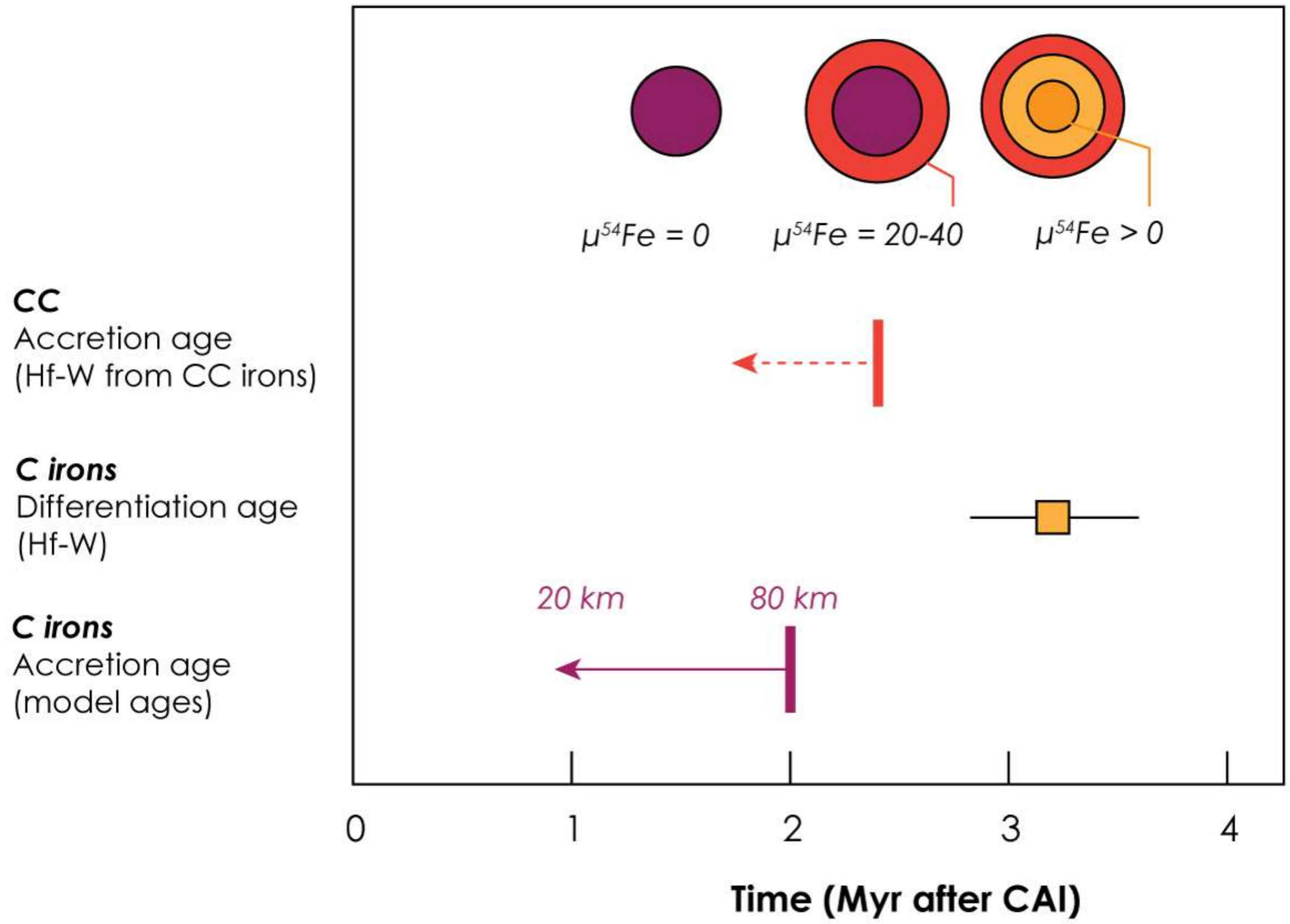

**Fig. 6: Accretion and differentiation timescales of outer disk planetesimals**, with model accretion ages for C irons (depending on their size)[29], their differentiation ages from Hf-W dating[30] and the modelled accretion ages of CCs[31]. The top schematic reveals the changes in Fe isotope composition when starting with CI-like bodies. Recent nucleosynthetic Fe and Ni isotope analyses of ungrouped carbonaceous iron meteorites, together with carbonaceous chondrites, have revealed a mixing trend between an ODD-like and a CI-like isotopic endmember [32]. Notably, in contrast to other carbonaceous chondrite groups, CI chondrites lack isotopically related iron meteorite counterparts. This observation has been interpreted to reflect the late accretion of CI-like parent bodies—after the decay of $^{26}$Al had largely ceased—thereby preventing differentiation, whereas the parent bodies of carbonaceous irons accreted early. In this interpretation, the data implies a late-stage addition of CI-like material to an originally ODD-like disk. Here, we suggest an alternative explanation for this apparent discrepancy, taking into account the cosmochemical timeline of planetary accretion and differentiation in the outer solar system. While CC iron meteorite parent bodies likely accreted early (i.e., <2 Myr after CAI formation) [29], their Hf-W metal-silicate differentiation ages cluster around ~3.2 Myr [30]. This places the accretion age of carbonaceous chondrites in between, at ~2.5 Myr after CAI formation [31]. Assuming that these CC iron parent bodies initially accreted from CI-like materials, their compositions were likely modified by other CC-like materials. Such heterogeneous accretion has been postulated for CM and CR chondrites, where the anomalous chondrites Bells and NWA 14674 have been put forward as examples [5,9]. The assumption would be that this late addition to the initial parent body would also melt and contribute to the core composition. Consequently, the isotope composition of CC irons is unlikely to be CI-like, but always a mixture of CI and ODD endmembers.

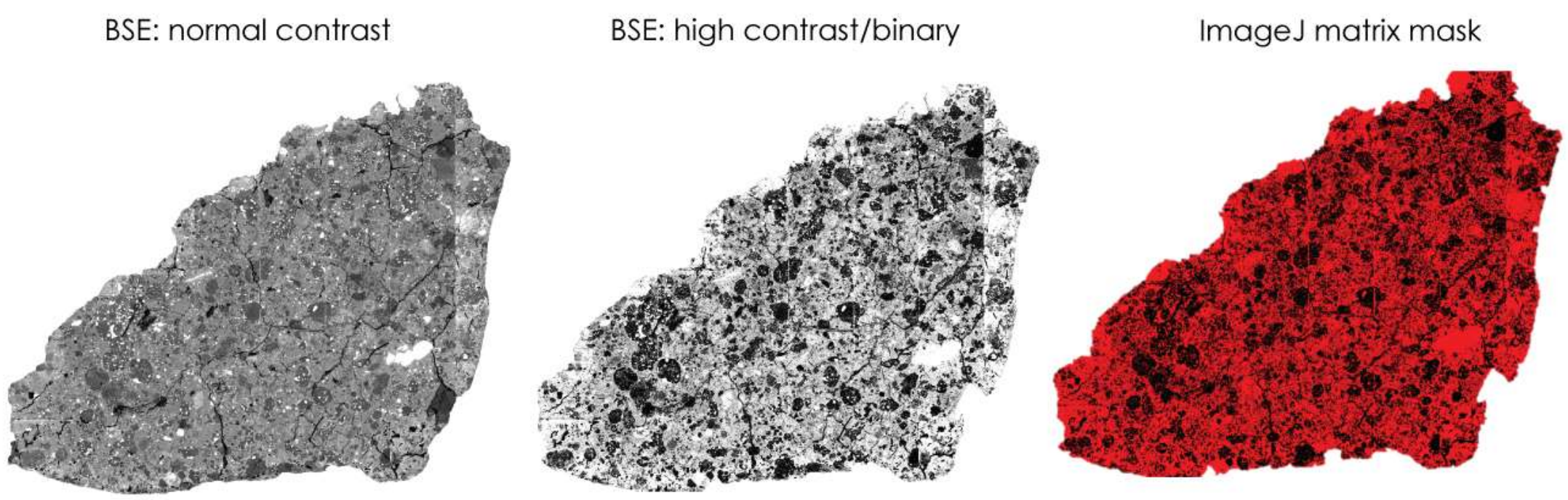

**Fig. 7: Process of determining the chondrule to matrix ratio of chondritic sections** in this study. Obtained BSE images are modified to high contrast binaries, where chondrules appear dark and matrix light. A threshold is then used in Fiji (ImageJ) to mask the matrix region and calculate the percent area. NWA 16476 is used as an example here. Note that this process is not perfect, since the matrix fraction will include type II chondrules (<5 vol.%) and metals/sulfides within chondrules. To correct for the latter, we included another threshold to calculate the high-density phase volume percentage, which was subtracted from the matrix fraction. In some cases where the contrast between chondrules and matrix was very low, we took values from literature.

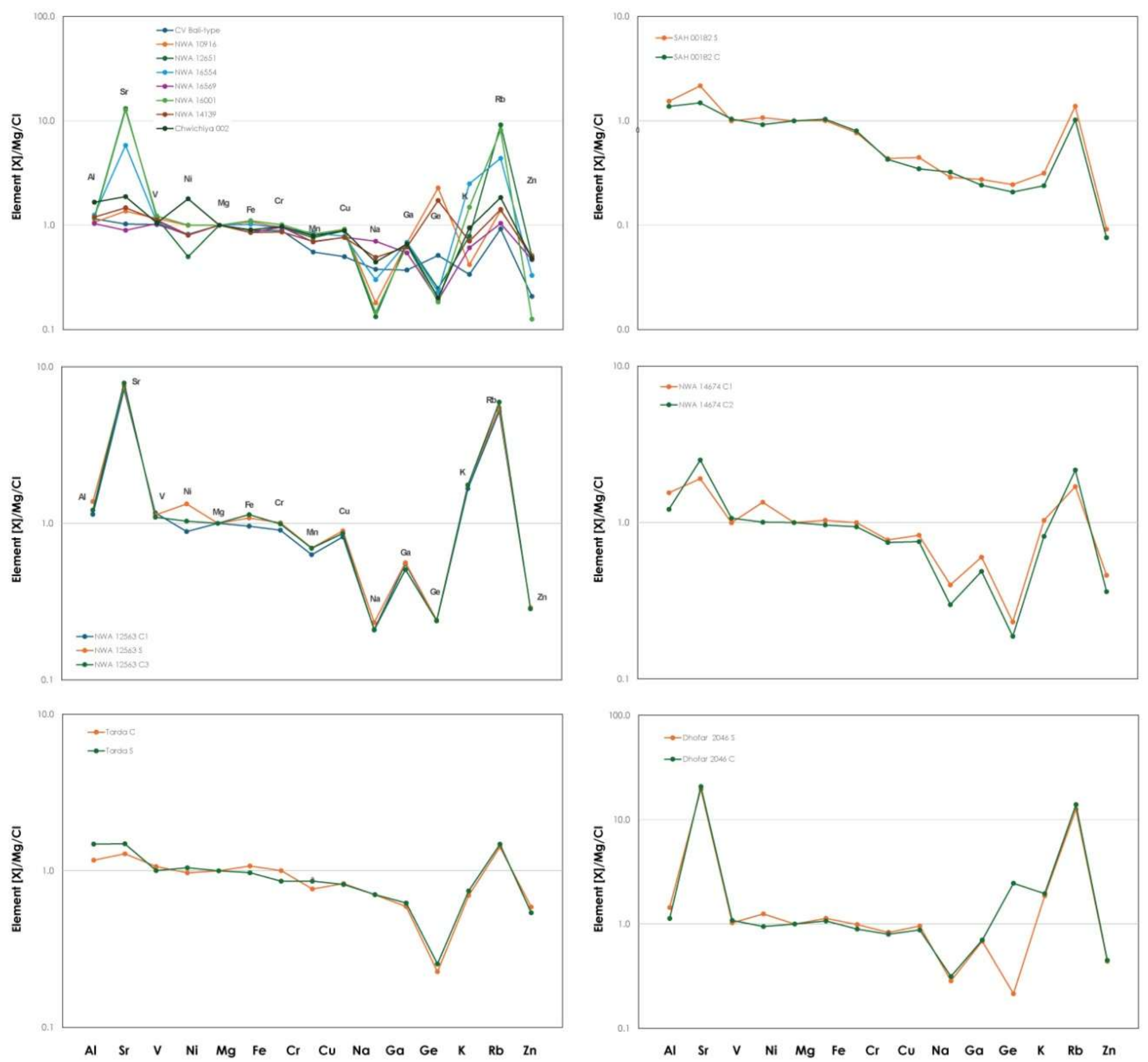

**Fig. 8: Major and minor element data for bulk ungrouped and grouped chondrites** from iCAP ICPMS in order of their relative volatility (see Methods). We show examples of saw dust (with fusion crust) and crushed (without fusion crust) aliquots. For the overall chemical composition, the addition of fusion crust to the sample seems to be negligible for handpieces >2 cm in diameter. Some chondrites show large enrichments of Sr and Rb relative to CI chondrites (<20 × CI for Sr), which we attribute to terrestrial weathering in a dessert environment. This secondary alteration may also influence the Na and K concentrations of the samples.

| | | $\mu^{54}$Fe | 2SE | $\mu^{54}$Cr | 2SE | $\mu^{30}$Si | 2SE | $\mu^{26}$Mg* | 2SE | $^{27}$Al/$^{24}$Mg | |
|---|---|---|---|---|---|---|---|---|---|---|---|
| Earth | | 2 | 2 | 10.0 | 13.0 | 1.20 | 1.50 | 0.0 | 3 | 0.085 | * |
| Mars | | 6.5 | 1.5 | -20.5 | 3 | -5.80 | 3.00 | -4.7 | 1.9 | 0.085 | * |
| HED | | 11.7 | 2.4 | -73 | 8 | -9.50 | 4.00 | -4.9 | 2 | 0.085 | * |
| Ureilite | | 13.9 | 2.8 | -90 | 15 | -9.50 | 2.80 | -8.2 | 2 | 0.085 | * |
| EC | | 6.4 | 0.7 | 2.0 | 11.0 | 10.97 | 2.58 | 1.1 | 2.2 | 0.085 | * |
| OC | | 10.5 | 2.6 | -39.0 | 9.0 | 7.43 | 1.03 | -4.0 | 2.6 | 0.078 | |
| R | NWA 753 | 6.4 | 0.8 | -7.0 | 3.0 | 1.10 | 3.10 | -1.3 | 2.3 | 0.085 | |
| CR | | 28.8 | 4.4 | 128.0 | 23.0 | 9.86 | 4.12 | -4.4 | 3.9 | 0.120 | |
| CV | NWA 2364 ($CV_{OX,A}$) | 22.1 | 3.6 | 89.0 | 30.0 | 16.40 | 2.00 | 2.6 | 1.9 | 0.129 | |
| CM | Murchison | 26.7 | 5.4 | 92.0 | 26.0 | 16.80 | 7.10 | 4.9 | 2.5 | 0.111 | |
| CO | NWA1231 | 16.6 | 7.2 | 90.0 | 43.0 | 20.30 | 5.30 | 1.6 | 4 | 0.101 | |
| CK | | 26 | 6.1 | 51.0 | 15.0 | 20.35 | 3.10 | 7.0 | 1.8 | 0.111 | |
| CI | | 3 | 2 | 156.0 | 5.0 | 32.44 | 2.41 | 4.5 | 0.4 | 0.096 | |
| ODDs | | 33 | 4 | 125.0 | 9.0 | 4.70 | 1.80 | -16.0 | 1.5 | 0.098 | |
| Tagish Lake | C2-ung | 16 | 5 | 119.0 | 15.0 | 27.0 | 7.4 | 1.2 | 2.8 | 0.104 | |
| Tarda | C2-ung | 23.9 | 2.5 | 137.0 | 16.0 | 24.9 | 2.8 | 2.7 | 2.1 | 0.109 | |
| Chwichiya 002 | CT3 | 28 | 1.6 | 90.7 | 8.3 | 20.4 | 5.4 | 3.4 | 1.5 | 0.110 | |
| NWA 14139 | CT3 | 23 | 4.9 | 104.1 | 8.7 | 18.8 | 6.4 | 9.2 | 1.8 | 0.121 | |
| NWA14051 | CT3 | 24.1 | 6.2 | 87.2 | 5.2 | 8.6 | 5.1 | 0.7 | 1.4 | 0.103 | |
| NWA16001 | CT3 | 24.4 | 6.4 | 91.3 | 9.6 | 24.9 | 5.2 | 4.4 | 1.9 | 0.108 | |
| NWA 16569 (1) | CT3 | 20.6 | 5.2 | 105.1 | 3.3 | | | 1.1 | 1.8 | 0.102 | |
| NWA 16569 (2) | CT3 | 18.6 | 1.4 | 74 | 7.6 | 22.5 | 3.1 | 3.2 | 1.2 | 0.113 | |
| NWA 16559 | CT3 | 23.0 | 2.9 | 97.8 | 7.8 | 27.9 | 3.1 | 0.1 | 1.7 | 0.108 | |
| NWA 16554 (1) | CM-an | 22.6 | 4.3 | 79.1 | 6.3 | 18.1 | 6.9 | 6.3 | 1.3 | 0.115 | |
| NWA 16554 (2) | CM-an | 18.0 | 2.5 | 104.8 | 6.3 | 33.8 | 3 | 6.4 | 2.2 | 0.112 | |

| | | $\mu^{54}$Fe | 2SE | $\mu^{54}$Cr | 2SE | $\mu^{30}$Si | 2SE | $\mu^{26}$Mg* | 2SE | $^{27}$Al/$^{24}$Mg |
|---|---|---|---|---|---|---|---|---|---|---|
| NWA 16000 | CM-an | 21.6 | 2.5 | 119.9 | 6.5 | 30.6 | 2.8 | 1.9 | 2.5 | 0.111 |
| Dhofar 2046 | CM-an | 23.7 | 3.3 | 121.3 | 7.1 | 19.4 | 6.0 | 2.2 | 2.6 | 0.104 |
| NWA 12563 | CM-an | 9.5 | 5.2 | 89.9 | 5.1 | 38.6 | 7.3 | 9.0 | 1.5 | 0.115 |
| | | | | | | | | | | |
| NWA 13400 | CL3.9 | 18.6 | 3.4 | 65.2 | 3.5 | 23.1 | 5.1 | 4.7 | 1.4 | 0.114 |
| NWA 033 | CL4 | 25.9 | 7.1 | 64.6 | 7.2 | 17.4 | 2.4 | 1.1 | 1.6 | 0.120 |
| | | | | | | | | | | |
| SAH 00182 | C3-ung | 24.7 | 5.3 | 72.4 | 5.2 | 8.2 | 1.1 | 5.0 | 1 | 0.123 |
| NWA 14674 | CR2-an | 31.2 | 3.3 | 118.5 | 5.6 | 10.4 | 6.9 | 3.4 | 1.6 | 0.118 |
| Ankazomena | CVox3 | 29.5 | 8 | 97.9 | 5.7 | 7.3 | 4.9 | 2 | 2.2 | 0.110 |
| | | | | | | | | | | |
| NWA 16476 | CO3 | 22.4 | 2.5 | 131.4 | 9.7 | 32.2 | 3.7 | 8 | 2.4 | 0.116 |
| NWA 11085 | CO3.0 | 15.0 | 7.9 | | | 24.7 | 3.9 | 7.3 | 2 | 0.114 |
| NWA 10916 | CO3.0 | 25.8 | 5.6 | | | 21.7 | 7.4 | -1.2 | 1.4 | 0.100 |
| NWA12651 | CM | 23 | 5.1 | 100.8 | 4.1 | 14.6 | 5.6 | 1.6 | 1.4 | 0.114 |

**Table 1:** The average mass-independent Fe, Cr, Si, and Mg isotope compositions and $^{27}$Al/$^{24}$Mg ratios of bulk ungrouped carbonaceous chondrites and their standard errors. Grouped chondrite and achondrite isotope values are from literature[9–12,20,33,34] and their $^{27}$Al/$^{24}$Mg ratios are taken from [20,35]. * The initial $^{27}$Al/$^{24}$Mg ratios of Earth, Mars, Vesta and Ureilites are assumed to be like inner Solar System chondrites. Individual sample analyses and mass-dependent isotope data can be found in *raw data table R2*. We also show the results for the three different methods to correct the obtained $\mu^{26}$Mg* values back to the solar $^{27}$Al/$^{24}$Mg ratio of 0.096[22] (see supplementary text, ***Fig. 4-5***).

| | | Method 1 (canonical $^{26}$Al) | | Method 2 (initial $\mu^{26}$Mg* = -38) | Method 3 (initial $\mu^{26}$Mg* = -15.9) |
|---|---|---|---|---|---|
| | | **Initial $\mu^{26}$Mg*** | **$\mu^{26}$Mg*** | **$\mu^{26}$Mg*** | **$\mu^{26}$Mg*** |
| Earth | | -32.6 | 4.2 | 4.9 | 2.1 |
| Mars | | -37.3 | -0.5 | -0.4 | -3.3 |
| HED | | -37.5 | -0.7 | -0.6 | -3.5 |
| Ureilite | | -40.8 | -4.0 | -4.3 | -7.2 |
| EC | | -31.6 | 5.2 | 6.1 | 3.3 |
| OC | | -34.0 | 2.9 | 3.8 | -1.3 |
| R | NWA 753 | -33.9 | 2.9 | 3.4 | 0.6 |
| CR | | -50.6 | -13.7 | -11.2 | -6.7 |
| CV | NWA 2364 | -46.9 | -10.0 | -7.8 | -2.1 |
| CM | Murchison | -37.7 | -0.9 | -0.9 | 2.1 |
| CO | NWA1231 | -37.3 | -0.4 | -0.5 | 0.7 |
| CK | | -35.6 | 1.2 | 0.9 | 3.9 |
| CI | | -32.4 | 4.5 | 4.5 | 4.5 |
| ODDs | | -53.6 | -16.8 | -16.4 | -16.0 |
| Tagish Lake | C2-ung | -38.7 | -1.9 | -1.8 | -0.1 |
| Tarda | C2-ung | -39.2 | -2.3 | -2.2 | 0.5 |
| Chwichiya 002 | CT3 | -38.8 | -2.0 | -1.9 | 0.9 |
| NWA 14139 | CT3 | -37.1 | -0.2 | -0.4 | 4.1 |
| NWA14051 | CT3 | -38.8 | -1.9 | -1.9 | -0.4 |
| NWA16001 | CT3 | -37.1 | -0.2 | -0.3 | 2.1 |
| NWA 16569 (1) | CT3 | -38.1 | -1.2 | -1.2 | 0.1 |
| NWA 16569 (2) | CT3 | -40.2 | -3.3 | -3.0 | 0.3 |
| NWA 16559 | CT3 | -41.4 | -4.5 | -4.1 | -1.7 |
| NWA 16554 (1) | CM-an | -37.9 | -1.0 | -1.0 | 2.6 |
| NWA 16554 (2) | CM-an | -36.6 | 0.3 | 0.1 | 3.2 |
| NWA 16000 | CM-an | -40.7 | -3.9 | -3.5 | -0.5 |
| Dhofar 2046 | CM-an | -37.7 | -0.9 | -0.9 | 0.8 |
| NWA 12563 | CM-an | -35.2 | 1.7 | 1.2 | 4.9 |
| | | | | | |
| NWA 13400 | CL3.9 | -39.1 | -2.2 | -2.0 | 1.4 |
| NWA 033 | CL4 | -45.0 | -8.1 | -6.7 | -2.3 |
| | | | | | |
| SAH 00182 | C3-ung | -42.2 | -5.4 | -4.4 | 0.4 |
| NWA 14674 | CR2-an | -41.9 | -5.0 | -4.3 | -0.2 |
| Ankazomena | CVox3 | -40.2 | -3.4 | -3.1 | -0.3 |
| | | | | | |
| NWA 16476 | CO3 | -36.5 | 0.3 | 0.1 | 3.9 |
| NWA 11085 | CO3.0 | -36.5 | 0.4 | 0.1 | 3.6 |
| NWA 10916 | CO3.0 | -39.8 | -2.9 | -2.8 | -1.9 |
| NWA12651 | CM | -42.2 | -5.3 | -4.7 | -1.2 |

**Table 2: The recalculated $\mu^{26}Mg^{*}$ values of anomalous carbonaceous chondrites** from supplementary table 1, based on analyzed $^{27}Al/^{24}Mg$ ratios and assumptions from three different correction methods described in supplementary figures 4 and 5.